\documentclass[a4paper,fleqn,usenatbib]{mnras}

\usepackage{newtxtext,newtxmath}
\usepackage{graphicx,subcaption,chngpage,calc}
\usepackage{graphicx}
\usepackage{lscape}
\usepackage{ltablex}
\usepackage{appendix}
\usepackage{float}
\usepackage{placeins}
\usepackage{afterpage}
\restylefloat{table}

\usepackage[T1]{fontenc}
\usepackage{ae,aecompl}

\usepackage{graphicx}	% Including figure files
\usepackage{amsmath}	% Advanced maths commands
\usepackage{amssymb}	% Extra maths symbols
\usepackage{multirow}
\usepackage{float}
\title[The disc--like host galaxies of RL NLSy1s]{The disc--like host galaxies of radio loud Narrow-Line Seyfert 1s}

\author[A. Olgu\'in--Iglesias]{Alejandro Olgu\'in--Iglesias,$^{1}$\thanks{E-mail: alejandroolguiniglesias@gmail.com}
Jari Kotilainen,$^{2,3}$
Vahram Chavushyan$^{1}$
\\
$^{1}$Instituto Nacional de Astrof\'isica, \'Optica y Electr\'onica, Luis E. Erro 1, Tonantzintla, Puebla, M\'exico,
C.P. 72840\\
$^{2}$ Finnish Centre for Astronomy with ESO (FINCA), Vesilinnantie 5, FI-20014 University of Turku, Finland\\
$^{3}$ Department of Physics and Astronomy, Vesilinnantie 5, FI-20014 University of Turku, Finland
}

\date{Accepted XXX. Received YYY; in original form ZZZ}

\pubyear{2015}

\begin{document}
\label{firstpage}
\pagerange{\pageref{firstpage}--\pageref{lastpage}}
\maketitle

\begin{abstract}
Until recently, relativistic jets were ubiquitously found to be launched from giant elliptical galaxies. However, the detection by the Fermi--LAT of $\gamma-$ray emission from radio--loud narrow--line Seyfert 1 (RL--NLSy1) galaxies raised doubts on this relation. Here, we morphologically characterize a sample of 29 RL--NLSy1s (including 12 $\gamma-$emitters, $\gamma-$NLSy1s) in order to find clues on the conditions needed by AGN to produce relativistic jets. We use deep near-infrared images from the Nordic Optical Telescope and the ESO VLT to analyze the surface brightness distribution of the galaxies in the sample. We detected 72\% of the hosts (24\% classified as $\gamma-$NLSy1s). Although we cannot rule out that some RL--NLSy1s are hosted by dispersion supported systems, our findings strongly indicate that RL--NLSy1s hosts are preferentially disc galaxies. 52\% of the resolved hosts (77\% non $\gamma-$emitters and 20\% $\gamma-$emitters) show bars with morphological properties (long and weak) consistent with models that promote gas inflows, which might trigger nuclear activity. The extremely red bulges of the $\gamma-$NLSy1s, and features that suggest minor mergers in 75\% of their hosts, might hint to the necessary conditions for $\gamma-$rays to be produced. Among the features that suggest mergers in our sample, we find 6 galaxies that show offset stellar bulges with respect to their AGN. When we plot the nuclear versus the bulge magnitude, RL--NLSy1s locate in the low--luminosity end of flat spectrum radio quasars (FSRQs), suggesting a similar accretion mode between these two AGN types. 
\end{abstract}


%%%%%%%%%%%%%%%%%%%%%%%%%%%%%%%%%%%%%%%%%%%%%%%%%%

%%%%%%%%%%%%%%%%% BODY OF PAPER %%%%%%%%%%%%%%%%%%

\section{Introduction}\label{sec:intro}

 Because of the tight empirical relations observed between the black hole mass and different properties of its 
 host galaxy bulge \citep{Magorrian_1998,gebhardt_2000, ferrarese_merrit_2000,tremaine_2002,Gultekin_2009}, 
 it is now widely accepted that there is a strong connection between the supermassive black holes (SMBHs) 
 and their host galaxies. 
 The link between the black hole and its host galaxy is thought to be the AGN activity, 
 through feedback processes \citep[either positive or negative; for a review see][]{Fabian_2012,Heckman_2014}. 
 If this is the case, then, we can assume that the more powerful the AGN, the stronger the influence on its host galaxy. 
 In fact, a study by \citealt{olguin_2016} (performed on $\mathrm{0.3<z<1.0}$, strongly beamed AGN, whose 
 relativistic jets point towards the Earth, i.e. blazars) suggests that the AGN 
 impact their hosts (either by suppression or triggering of star formation) in a magnitude that is proportional 
 to the jet power.\smallskip
 
 Until recently, powerful relativistic radio jets were virtually only found to be hosted in elliptical galaxies 
 \citep[e.g.][]{stickel_1991,Kotilainen_1998bllacs,kotilainen_1998,scarpa_2000,urry_2000,kotilainen_2005,
 Hyvonen_2007,olguin_2016}, which helped develop ideas on how jets, supermassive black holes 
 and their host galaxies evolve. However, recently, a few studies report on blazar--like disc hosts, 
 that is to say, with fully developed relativistic jets, capable of emitting $\gamma-$ray 
 photons \citep[][and probably \citealt{leontavares_2014}] 
 {kotilainen_2016,olguin_2017}. These blazar--like disc galaxies, constitute a peculiar type of 
 AGN known as Narrow--line Seyfert 1s (NLSy1s), characterized by narrower Balmer lines
 full width at half maximum ($\mathrm{FWHM(H\beta)<2000\ km\ s^{-1}}$) than in normal Seyferts, 
 flux ratios $\mathrm{[OIII]/ H\beta <3}$, strong optical FeII lines (FeII bump) 
 and a soft X-ray excess \citep{,osterbrok_pogge_1985,orban_2011}. Based on the full width at half maximum 
 (FWHM) of their Broad Line Region (BLR) lines and the continuum luminosity \citep{kaspi_2000}, their central black 
 holes masses ($\mathrm{M_{BH}}$) are estimated to range from $\mathrm{\sim 10^6M_\odot}$ to $\mathrm{\sim10^7M_\odot}$ \citep{mathur_2012} and 
 thus, their accretion rates are thought to be close to the Eddington limit. A small fraction 
 has been found to be radio loud  \citep[RL, 7\%,][]{komossa_2006} and among them, a smaller fraction (so far, 15 galaxies), 
 are the above-mentioned $\gamma-$ray emitters NLSy1s 
 (hereafter, $\gamma-$NLSy1s).\smallskip

 \begin{table*}
 \centering
  \begin{minipage}{175mm}
  \caption{Main properties of the RL--NLSy1s analysed in this work and observations log.}
 \label{table:sample}
 \resizebox{1.0\textwidth}{!}{%
   \begin{tabular}{@{}cccccccccl@{}}
   \hline
 \multirow{2}{*}{Source}	&	\multirow{2}{*}{Name}	&	\multirow{2}{*}{z}	&	\multirow{2}{*}{RA}	&	\multirow{2}{*}{DEC}	&	\multirow{2}{*}{UT Date} &	Scale 	&	Seeing	&	Exposure time \\	
 	&	 	&		&		&		&		&	kpc/''	&	(arcsec)&	(s)           \\
 \multicolumn{1}{c}{(1)}&	\multicolumn{1}{c}{(2)} 	& (3)	 &	(4)	& (5)		&	(6)	&	(7) &	(8)	&	\multicolumn{1}{c}{(9)}	  		 	\\	
 \hline
 0321$+$340$^\gamma$	&	1H 0323$+$342   	&	0.061	&	03:24:41.1	&	$+$34:10:46.0		&	23-Jan-13	&	1.177	&		1.00/1.00	&	1800/1800 \\
 0846$+$513$^\gamma$	&	SBS 0846+513    	&	0.580	&	08:49:58.0	&	$+$51:08:29.0		&	14-Feb-16	&	6.579	&		0.65/0.67	&	3000/2700 \\
 0929+533$^\gamma$	&	J093241+53063		&	0.590	&	09:32:41.1	& 	+53:06:33.3		& 	30-Mar-18	&	6.633	&		-/0.77		& 	-/4356 \\
 0948$+$002$^\gamma$	&	PMN J0948$+$0022	&	0.585	&	09:48:57.3	&	$+$00:22:26.0		&	21-Feb-14	&	6.606	&		0.73/0.77	&	4950/4890 \\
 0955+32$^\gamma$	&	J095820+322401		&	0.530	&	09:58:20.9	&	$+$32:24:01.6		&	31-Mar-18	&	6.293	&		-/0.78		&-/3600	\\
 1102+2239		&	FBQS J1102+2239		&	0.453	&	11:02:23.4	&	$+$22:39:20.7		&	14-Feb-14	&	5.781	&		-/0.79		&-/2700\\
 1159$-$011		&	IRAS 11598$-$0112	&	0.150	&	12:02:26.8	&	$-$01:29:15.0		&	13-Feb-16	&	2.614	&		0.79/0.82	&   2280/900  \\
 1200$-$004		&	RXSJ12002$-$0046	&	0.179	&	12:00:14.1	&	$-$00:46:39.0		&	14-Feb-16	&	3.021	&		0.64/0.67	&	1080/960  \\
 1217$+$654		&	J12176$+$6546		&	0.307	&	12:17:40.4	&	$+$65:46:50.0		&	13-Feb-16	&	4.526	&		0.94/0.79	&	2160/2160 \\
 1219$-$044$^\gamma$	&	4C$+$04.42      	&	0.996	&	12:22:22.5	&	$+$04:13:16.0		&	29-Mar-14	&	8.001	&		0.80/0.84	&	3650/2160 \\
 1227$+$321		&	RXSJ12278$+$3215	&	0.137	&	12:27:49.2	&	$+$32:14:59.0		&	14-Feb-16	&	2.423	&		0.70/0.67	&	930/930	  \\
 1246+0238$^\gamma$	&	SDSS J124634.65+023809	& 0.362		& 12:46:34.6		&02:38:09.1			&	12-Aug-17	& 	5.048	&		-/0.80		&	-/3480\\
 1337$+$600		&	J13374$+$6005  		&	0.234	&	13:37:24.4	&	$+$60:05:41.0		&	13-Feb-16	&	3.722	&		0.73/0.72	&	2160/2160 \\
 1403$+$022		&	J14033$+$0222 		&	0.250	&	14:03:22.1	&	$+$02:22:33.0		&	14-Feb-16	&	3.910	&		0.66/0.59	&	1380/1080 \\
 1421+3855$^\gamma$	&J142106+385522$^\gamma$	&	0.490	&	14:21:06.0	&	+38:55:22.5		&	14-Mar-18	&	6.038 	&		-/0.75		&-/2754\\
 1441$-$476$^\gamma$	&	B3 1441$+$476   	&	0.705	&	14:43:18.5	&	$+$47:25:57.0		&	24-Apr-16	&	7.166	&		-/0.80		&	-/2700 \\
 1450+591		&	J14506+5919		&	0.202	&	14:50:42.0	&	$+$59:19:37		&	22-May-14	&	3.325  	&		0.82/0.82	&	900/1020\\	
 1502$+$036$^{\gamma \ *}$ 	&	PKS 1502$+$036  	&	0.409	&	15:05:06.5	&	$+$03:26:31.0		&	04-Apr-13	&	5.446	&		0.80/0.82	&	920/280   \\
 1517$+$520		&	SBS 1517$+$520 		&	0.371	&	15:18:32.9	&	$+$51:54:57.0		&	14-Feb-16	&	5.128	&		0.63/0.67	&	1140/960  \\
 1546$+$353		&	B2 1546$+$35A 		&	0.479	&	15:48:17.9	&	$+$35:11:28.0		&	13-Feb-16	&	5.964	&		0.68/0.74	&  2160/2340  \\
 1629$+$400		&	J16290$+$4007 		&	0.272	&	16:29:01.3	&	$+$40:08:00.0		&	13-Feb-16	&	4.157	&		0.63/0.62	&	2160/2160 \\
 1633$+$471		&	RXSJ16333$+$4718 	&	0.116	&	16:33:23.5	&	$+$47:19:00.0		&	14-Feb-16	&	2.101	&		0.62/0.63	&	1260/1050 \\
 1640$+$534		&	2E 1640$+$5345 	    	&	0.140	&	16:42:00.6	&	$+$53:39:51.0		&	14-Feb-16	&	2.468	&		0.60/0.64	&	1200/1080 \\
 1641$+$345		&	J16410$+$3454 	    	&	0.164	&	16:41:00.1	&	$+$34:54:52.0		&	14-Feb-16	&	2.814	&		0.62/0.70	&	1320/1800 \\
 1644$+$261$^\gamma$	&	FBQS J1644+2619 	&	0.145	&	16:44:42.5	&	$+$26:19:13.0		&	01-May-15	&	2.541	&		0.75/0.63	&	2550/2160 \\
 1702+457		&	B31702+457		&	0.060	&	17:03:30.3	&	$+$45:40:47.2		&	21-Jun-16	&	1.159	&		-/0.79		&	-/945 \\	
 1722$+$565		&	J17221$+$5654 	    	&	0.426	&	17:22:06.0	&	$+$56:54:51.0		&	13-Feb-16	&	5.579	&		0.62/0.60	&	2040/2040 \\
 2004$-$447$^{\gamma \ *}$	&	PKS 2004$-$447  	&	0.240	&	20:07:55.2	&	$-$44:34:44.0		&	16-Apr-13	&	3.793	&		0.40/0.45	&	600/110   \\
 2245$-$174		&	IRAS 22453$-$1744   	&	0.117	&	22:48:04.2	&	$-$17:28:30.0		&	14-Feb-16	&	2.116	&		1.44/-	 	&	1260/-      \\
 \hline
 \end{tabular}}\\
 
 \small {Columns: 
 (1) and (2) give the designation and name of the source;
 (3) the redshift of the object; (4) and (5) the J2000 right ascension and declination of the source; 
 (6) the observation date;
 (7) the target scale;
 (8) the seeing during the observation in J-- and Ks--band, respectively, and
 (9) the total exposure time for J-- and Ks--band, respectively.}\\
 $^{*}$ \scriptsize{Galaxies observed using the ISAAC on the ESO/VLT} \\
 $^\gamma$ \scriptsize{$\gamma-$ray emitting NLSy1 galaxies}
 \end{minipage}
 \end{table*}

 RL--NLSy1s (including $\gamma-$NLSy1s), are excellent 
 laboratories to study the mechanisms that make AGN able to launch and collimate fully developed 
 relativistic outflows at a likely early evolutionary stage. Thus, in this study, we characterize a 
 sample of  RL--NLSy1s (including 12 $\gamma-$NLSys detected so far) with the aim 
 of determining the properties of their host galaxies that could shed some light on the necessary conditions and mechanisms to 
 generate the relativistic jet phenomenon.\smallskip
 
 This paper is organized as follows. In Section \ref{sample},
 we present the sample and observations. In Section \ref{reduction} we explain the data 
 reduction and the methodology of the analysis. In Section \ref {results} we discuss the results and compare
 them with previous studies. Finally, in Section \ref{summary}, we summarize our findings.
 All quantitative values given in this paper are based on
 a cosmology with $\mathrm{H_0=70kms^{-1}Mpc^{-1}}$, $\mathrm{\Omega_m=0.3}$, and  $\mathrm{\Omega_{\Lambda} = 0.7}$.
 
 \section{Sample and observations} \label{sample}
 The initial sample consists of 12 $\gamma-$ray emitting NLSy1s 
 \citep{abdo_2009a,abdo_2009,foschini_2011,dammando_2012,dammando_2015,liao_2015,yao_2015,paliya_2018}. 
 Given that $\gamma-$NLSy1s are also radio loud (RL), we expanded this sample by imaging the host galaxies 
 of 17 radio--loud, but not $\gamma-$ray emitting NLSy1s, as a comparison sample. These galaxies are all 
 observable from the northern hemisphere, have redshifts $\mathrm{z<0.5}$
 and radio--loudness ($\mathrm{RL\equiv f_{\nu,4.85GHz}/f_{\nu,B}, RL>31}$). \smallskip 
 
 The observations were conducted using two different 
 telescopes, the Nordic Optical Telescope (NOT), 
 using the near--infrared camera NOTcam (pixel scale = $0.234 ''$/pixel and field-of-view FOV$=4'\times4'$) and the ESO very large telescope 
 (VLT), using its infrared spectrometer and array camera \citep[ISAAC, pixel scale =$0.148''$ and FOV = $152'' \times 152''$][]{Isaac}, depending on the declination 
 of each galaxy. On the other hand the RL-NLSy1s (but not $\gamma-$ray emitters) in the sample 
 were observed with the NOT telescope, between 2013 January 23 and 2016 February 14 using 
 the NOTcam.\smallskip
 
 As is usual in the near--infrared, all the targets in the sample were imaged using a jitter 
 procedure to obtain a set of offset frames with respect to the 
 initial position. Each target was observed in J-- and/or K--bands, during an average exposure time of $\mathrm{EXPTIME\approx1800s}$, 
 and an average seeing $\sim0.75$arcsec  (see table \ref{table:sample}). \smallskip
 
 \section{Data reduction and analysis}\label{reduction}
 \subsection{Data reduction}\label{calibration}
 The images of the galaxies in the sample observed with the NOT were reduced using the NOTCam script for IRAF 
 REDUCE \footnote{http://www.not.iac.es/instruments/notcam}. This script takes the consequtive dithered images, corrects for 
 flatfield, interpolates over bad pixels, and makes a sky template that is subtracted from each image. The 
 images are then registered based on interactively 
 selected stars (and RA/DEC header keywords) and combined to obtain the final reduced image. For the images observed with the ISAAC a similar procedure was followed. A flat frame was derived from the 
 twilight images and a sky image was obtained by median filtering the individual frames 
 in the stack. The individual frames were then aligned using bright stars as reference points in the field and 
 combined to produce the final reduced co--added image \citep[see][for more details]{olguin_2016}. The photometric 
 calibration was performed by using the field stars in our images with the magnitudes reported by 2MASS \citep{2mass}.
 
 \subsection{Photometric decomposition}
 The 2D light distribution of the reduced images is modeled using the image analysis algorithm GALFIT \citep{peng_2011}. 
 The different components of a galaxy are described using 
 different analytical functions. For bulges and possible bars in the host galaxies of the sample, 
 we use the S\'ersic profile, which functional form is:
 
 \begin{equation}
 \mathrm{\Sigma(r)=\Sigma_e exp\left[-\kappa\left(\left(\frac{r}{r_e}\right)^{1/n}-1\right)\right]}
 \label{eqn:sersic}
 \end{equation}
 
 where $\mathrm{\Sigma_e}$ is the surface brightness of the pixel in the effective radius ($\mathrm{r_e}$, radius where half of the total flux is concentrated). The parameter n (the S\'ersic index) is often 
 referred to as a concentration parameter and the variable $\mathrm{\kappa}$ is coupled to it.\smallskip 
 
 We also use the exponential function, since it describes well the radial behavior of galactic discs. 
 Although, the exponential function is a special case of the S\'ersic 
 profile (when $\mathrm{n=1}$), nomenclature--wise, we use it when the component to fit is a disc, 
 otherwise we use the S\'ersic profile with $\mathrm{n=1}$. Its functional form is:
 
 \begin{equation}
 \mathrm{\Sigma(r)=\Sigma_0exp\left(-\frac{r}{r_s}\right)}
 \label{eqn:disk}
 \end{equation}
  
 where $\mathrm{\Sigma(r)}$ is the surface brightness at a radius r, $\mathrm{\Sigma_0}$ is the surface brightness at the center of the target and $\mathrm{r_s}$ is the scale length of the disc.\smallskip
 
 The sky background is also modeled. We use a simple flat plane that can be tilted in x and y direction. Finally, the nuclear emission due to the powerful AGN of the galaxies 
 of the sample is fitted using a modeled point spread function (PSF). 
 
 \subsubsection{PSF modeling}
 
\begin{figure}
 \centering 
       \begin{tabular}{c} 
       \includegraphics[width=2.5in]{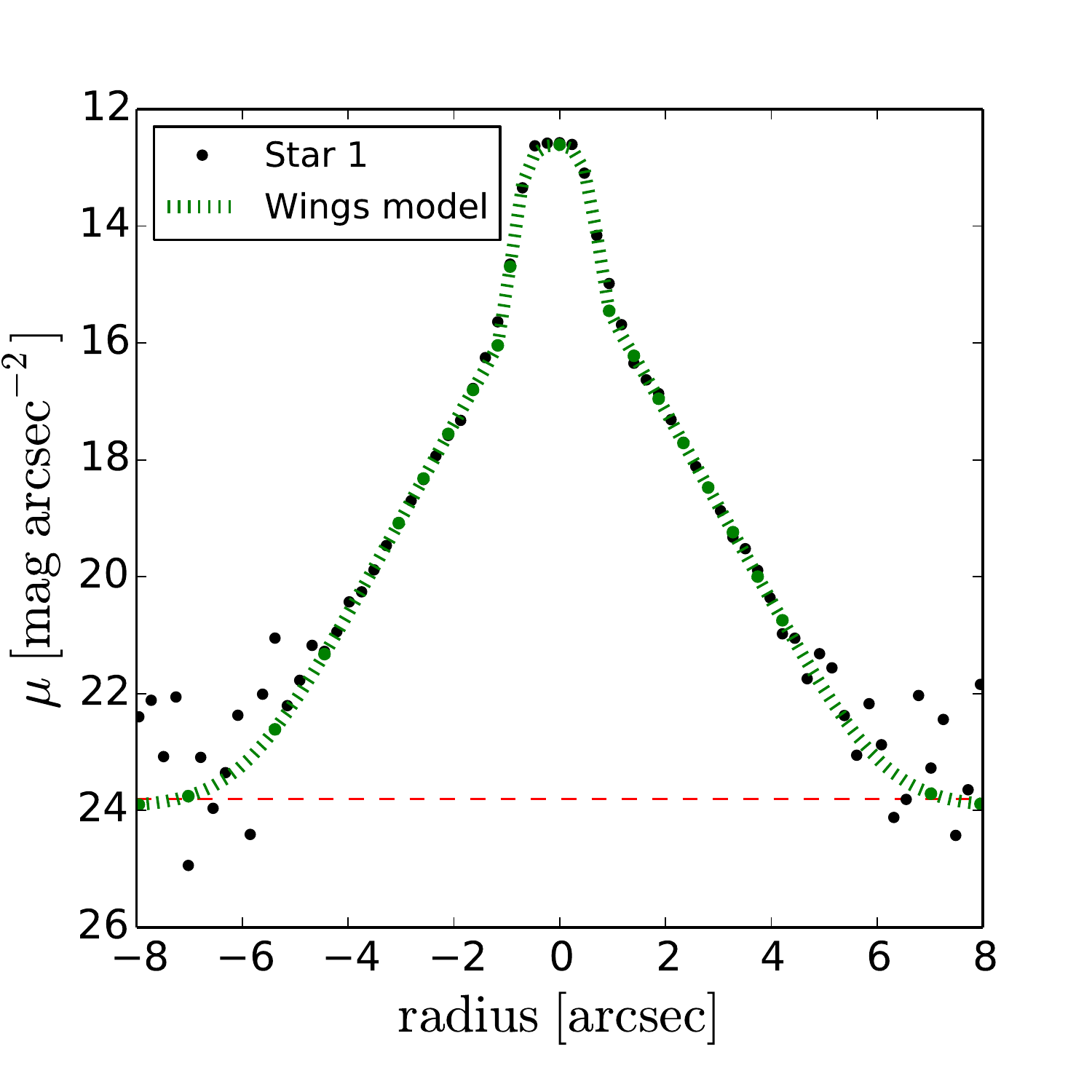}\\  
       \includegraphics[width=2.5in]{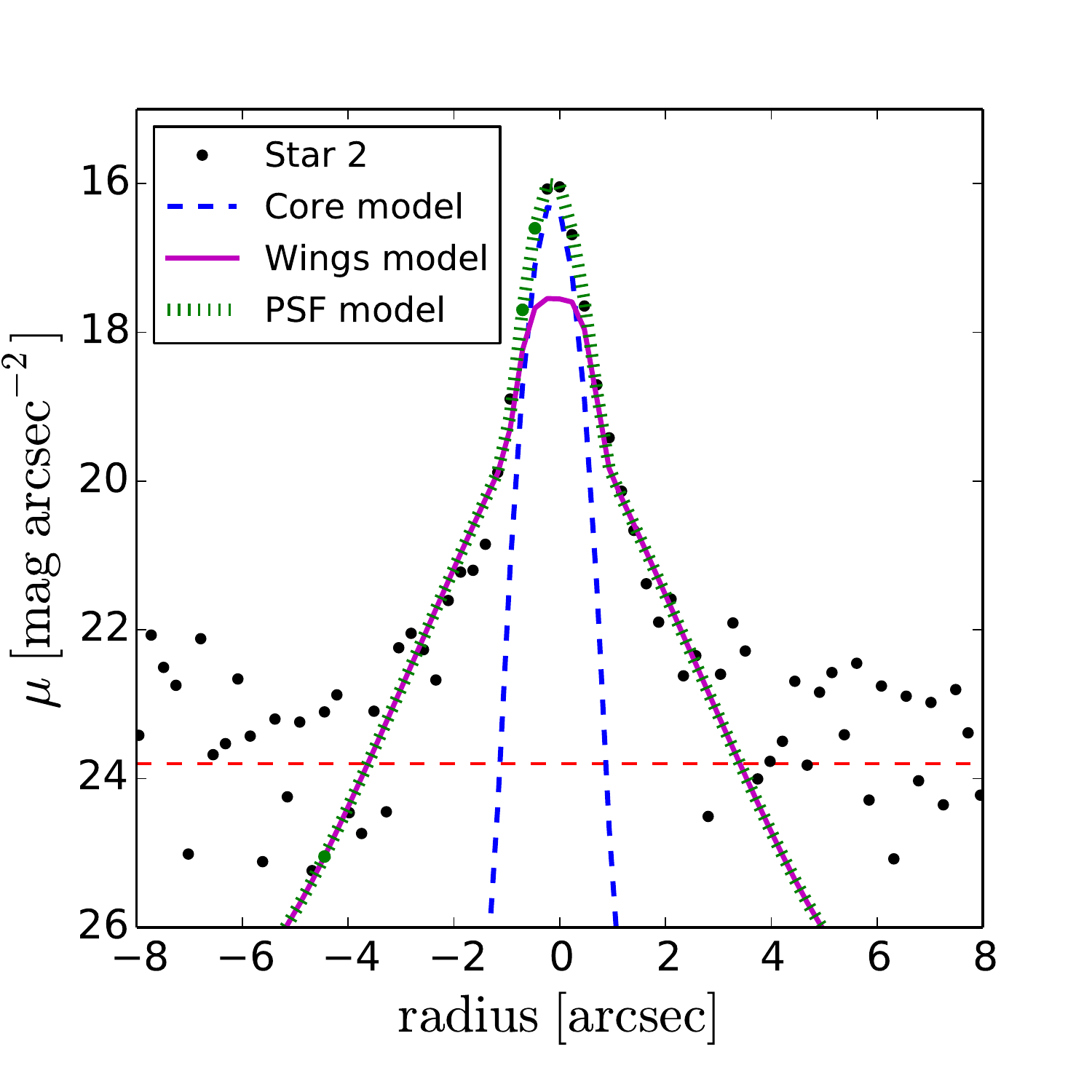}\\ 
       \end{tabular}
       \caption{PSF modeling procedure. In top panel, we show the PSF wings modeling 
       by fitting Gaussians and exponential functions to the surface brightness of a saturated star (Star 1).
       Since this star is saturated, there is no information on the PSF core.  
       In bottom panel, we show the PSF core modeling by fitting a Gaussian function to the surface 
       brightness of a non saturated star (star 2). As can be seen, one part of the wings in this 
       star is under the detection limit (red horizontal segmented line) and another part is noisy, 
       hence the need of the saturated star in order to derive the PSF wings.}\label{PSF}
 \end{figure}

 In most cases, the PSF modeling, only consists on subtracting a bright 
 \footnote{{\scriptsize we consider a star bright if it is brighter than 
 the target to fit. A PSF model made from a bright star can fit the wings of 
 the target (and beyond) and its nucleus. Otherwise, a faint star 
 (fainter that the target), will only be able to fit the nucleus and maybe part of the wings.}} 
 non--saturated star close to the target and removing its background. 
 However, it is not always possible to get a suitable star in the field of view (FOV). 
 In the case where only saturated or faint stars are found, 
 the following procedure is implemented:\smallskip
 
 First, we identify the stars in the FOV. Then, we select, preferentially, 
 the stars with no sources within $\sim$7'' radius and more than $\sim$10'' away 
 from the border of the FOV. The selected stars are centered in 50''$\times$50'' 
 boxes, where all extra sources are masked out by means of the segmentation 
 image process of SExtractor \citep{bertin_1996}. 
 The wings of the PSF are modeled by fitting a saturated star with a number of exponential 
 and Gaussian functions (top panel Figure \ref{PSF}). The core of the PSF is 
 modeled by using another star 
 (in this case, it is important not to be saturated) with Gaussian functions 
 and the previously generated wings model (bottom panel Figure \ref{PSF}). The magnitude difference between 
 the saturated and non--saturated stars is 
 important, since there must be an overlap in order to match the wings and 
 core models. The resultant model is tested by fitting random 
 stars in the FOV.\smallskip
 
In order to take into account the image PSF in the modeling of the galaxies, we convolved our PSF model with the analytical functions used in the fitting. The PSF model is also used to fit the nuclear component which, in the galaxies of this sample, is composed by the AGN.

 \subsubsection{Uncertainties}\label{errors}
 
 The uncertainties of the real functional form of a given galaxy component leads 
 to the errors in the galaxy models derived using the above-mentioned method. In order to 
 estimate this errors, we follow the procedure described by 
 \citet{olguin_2016,kotilainen_2016,olguin_2017} who identify three sources 
 of uncertainties; the PSF model, the sky 
 background and the zeropoint of photometric calibration.\smallskip 
 
 The uncertainties due to the PSF are accounted for by using different PSF models. 
 A number of PSF models can be derived using different 
 stars or different amount of Gaussians and exponentials. The uncertainties due 
 to the sky is derived by running several sky fits in separated regions 
 of 300 pixels $\times$ 300 pixels in the FOV. The zeropoint magnitude 
 depends on the star used to derive it, since they are estimated from 
 the magnitudes of several stars (see Section \ref{calibration}) then, we use the zeropoint 
 magnitude variations ($\sim\pm$0.1mag) as another source of uncertainties.\smallskip
 
 Using these variations, GALFIT is run several times. The resultant fits are used to make a 
 statistic where the best-fit value is the mean and the errors are 
 $\mathrm{\pm 1 \sigma}$ for every parameter of the galaxy model.
 
 \subsection{Simulations}\label{section:simulations}
In order to assure the suitability of the images to resolve galaxy bulges, we performed a set of simulations (Table \ref{table:sim}). The simulated galaxies have a nuclear component, represented by a star within the FOV. They also have a S\'ersic function, with $n=1$, which represents a bulge and an exponential function representing a disc. The background is taken from the same image as the star of the nuclear component. Every galaxy in the simulation has a different combination of parameters (bulge, disc and nuclear signal to noise ratios; bulge effective radius and seeing). The simulated galaxies are modeled in an identical way to the real galaxies in our sample in order to find whether the quality in our images is good enough to allow an acceptable subtraction of parameters. We found that the ability to properly retrieve the parameters of the bulges, depend both on the size of the bulge, with respect to the seeing, and on the brightness of the bulge with respect to the brightness of the other components. In this way, the parameters of faint bulges (with respect to the nucleus) might be properly retrieve provided that they are large enough (e.g. simulation 42). We note that, although the parameters of a bulge might not be properly retrieved, it might still be detected, although with not enough quality to retrieve its parameters. This means that in some simulations, we were able to properly characterize the nucleus and disc and also detect a residual (unable to be fitted) that we knew, beforehand, it was the bulge (e.g. simulation 1, 6 and 57). We focus on the parameters of bulge and nucleus, although all our simulations include a disc, with the intention of studying the effect of this component on the fittings. Most of the time the disc is not bright enough to hamper the bulge fit. However, the brightness of the bulge and the nucleus can often be high enough to make the modeling and even detection of the disc difficult (e.g. simulations 29, 30 and 75).\smallskip

\begin{figure}
 \centering
 \includegraphics[scale=0.35]{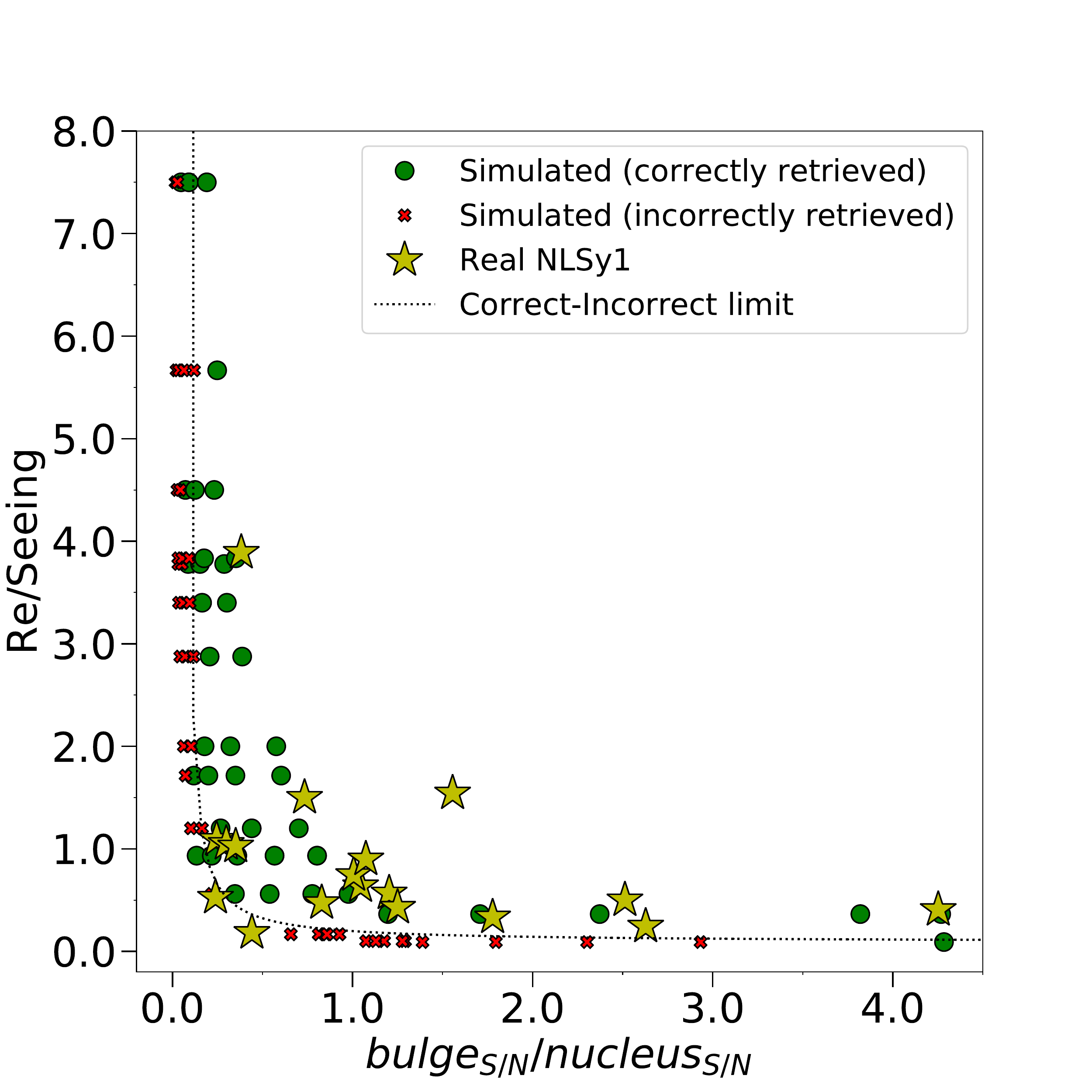}
 \caption{Ratio of the bulge over nucleus S/N versus bulge effective radius normalised to the seeing FWHM. 
 We see the location of the correctly retrieved simulated galaxies (green circles), 
 the incorrectly retrieved simulated galaxies (red crosses) and the real galaxies in our sample (yellow stars, including all 18 galaxies with K--band observations from the NOTcam and fitted in this study, and excluding 4 galaxies from previous studies and/or observed with the ESO/ISAAC, 6 unresolved galaxies and 1 observed only in J--band)
 A limit that divides the correctly from incorrectly retrieved simulated galaxies is represented by a segmented line.}
 \label{fig:simulations}
 \end{figure}

In figure \ref{fig:simulations} we show the set of simulations in a plot of the ratio between the bulge and nucleus S/N versus bulge effective radius normalized to the seeing FWHM. We use signal-to-noise ratios instead of magnitudes because in our simulations we take into account the seeing. The blurring caused by seeing makes the signal of point sources smaller. It also spreads out the light from extended sources, which reduces the measured signal-to-noise and the ability of telescopes to see detail but does not affect the derived magnitudes. Hence, the same source observed with different seeings have the same magnitude, but different signal-to-noise ratios, which is crucial in galaxy fitting. In our simulations, signal-to-noise ratios refer to the peak counts of the source divided by the background noise in the same region. We find the bulk of our sample in the region where the correctly retrieved simulated galaxies lie, which gives us confidence in the reliability of our analysis. We consider a model correct when the difference between a given parameter of the simulation and the same parameter of the model is less than the maximum error for such parameter in our analysis of the real galaxies sample (i.e. magnitude error $\mathrm{Mag_{err} \leq 0.50}$, bulge effective radius error $\mathrm{Re_{err}\leq 100\%}$ and S\'ersic index error $\mathrm{n_{err} \leq 1.35}$). Two galaxies in our sample lie outside that region (PMNJ0948+002 and J095820+322401), suggesting that those parameters are incorrectly retrieved, and hence left out of the analysis.

 \section{The host galaxies}\label{results}
In table \ref{table:galfit}, we show a summary of the NLSy1 host and nuclear properties derived from the two-dimensional surface brightness decomposition analysis. In the cases where the host galaxy is not detected we estimated upper limits for their magnitudes by simulating a galaxy \citep[following the method from][]{kotilainen_2007, olguin_2016}, assuming an average effective radius and S\'ersic index from the successfully detected hosts and then, increasing the magnitude until the component becomes detectable with a signal to noise ratio S/N=5, which by our own experience, is required in order to properly retrieve the structural parameters of the galaxy components.\smallskip 

In table \ref{table:galfit}, we summarize the nuclear and (disky--) bulge properties. However, in Appendix \ref{appendix_A}, the detailed fitting results are shown individually for every host in the sample. Out of the sample of 29 radio-loud NLSy1 host galaxies, 
 21 were successfully detected, 7 of which are $\gamma-$rays emitters and 14 are only radio-loud. For the sample, 
 we estimate an average J--band absolute nuclear and host magnitudes of $\mathrm{M(J){nuclear}=-23.8\pm 1.7}$ and $\mathrm{M(J){bulge}=-23.1\pm1.0}$ and an average K--band absolute nuclear and host magnitudes of $\mathrm{M(K){nuclear}=-25.6\pm1.4}$ and $\mathrm{M(K){bulge}=-24.8\pm1.1}$. The host luminosities for these galaxies are slightly fainter than elliptical galaxies hosting other types of radio--loud AGNs 
 \cite[e.g.][$\mathrm{M(K)_{FSRQs}=-26.2\pm0.9}$ and $\mathrm{M(K){BLLacs}=-25.6\pm 0.6}$]{olguin_2016} and rather similar to those of an $\mathrm{L^*}$ galaxy 
\citep[$\mathrm{M(K)=-24.8\pm 0.3}$][] {mobasher_1993}.\smallskip 

The average bulge effective radius for J--band and K--band respectively is $\mathrm{R_{eff}=1.9\pm1.3}$kpc and $\mathrm{R_{eff}=1.5\pm1.1}$kpc and the average S\'ersic index is $\mathrm{n=1.2\pm0.4}$ for J--band and $\mathrm{n=1.3\pm0.3}$ for K--band. The S\'ersic index can be used as an approximation to classify bulges \citep[either disky, $\mathrm{n<2}$ or classical, $\mathrm{n\ge2}$,][]{fisher_drory_2008}. According to the S\'ersic index derived from the 
 photometrical analysis (with the exception of J16410+3454 and J17221+5654, whose S\'ersic index might be $\mathrm{n>2}$ 
 given their uncertainties), all the NLSy1s hosts bulges are disc--like. However, it is well known that using the S\'ersic index to discern between classical from disc--like bulges can generate many misclassifications \citep{gadotti_2009}. To address this, we used the fact that disc--like and classical bulges are expected to be structurally different. Therefore, disc--like bulges should not follow the Kormendy relation 
 \citep[inverse relation between $\mathrm{\mu_e}$ and $\mathrm{R_e}$ found in elliptical galaxies and classical bulges;][]{Kormendy_1977}. In fact, \citet{gadotti_2009} finds that disc--like bulges lie below the Kormendy relation and hence, it can be used to distinguish them from classical bulges.

 \begin{figure}
 \centering
 \includegraphics[scale=0.4]{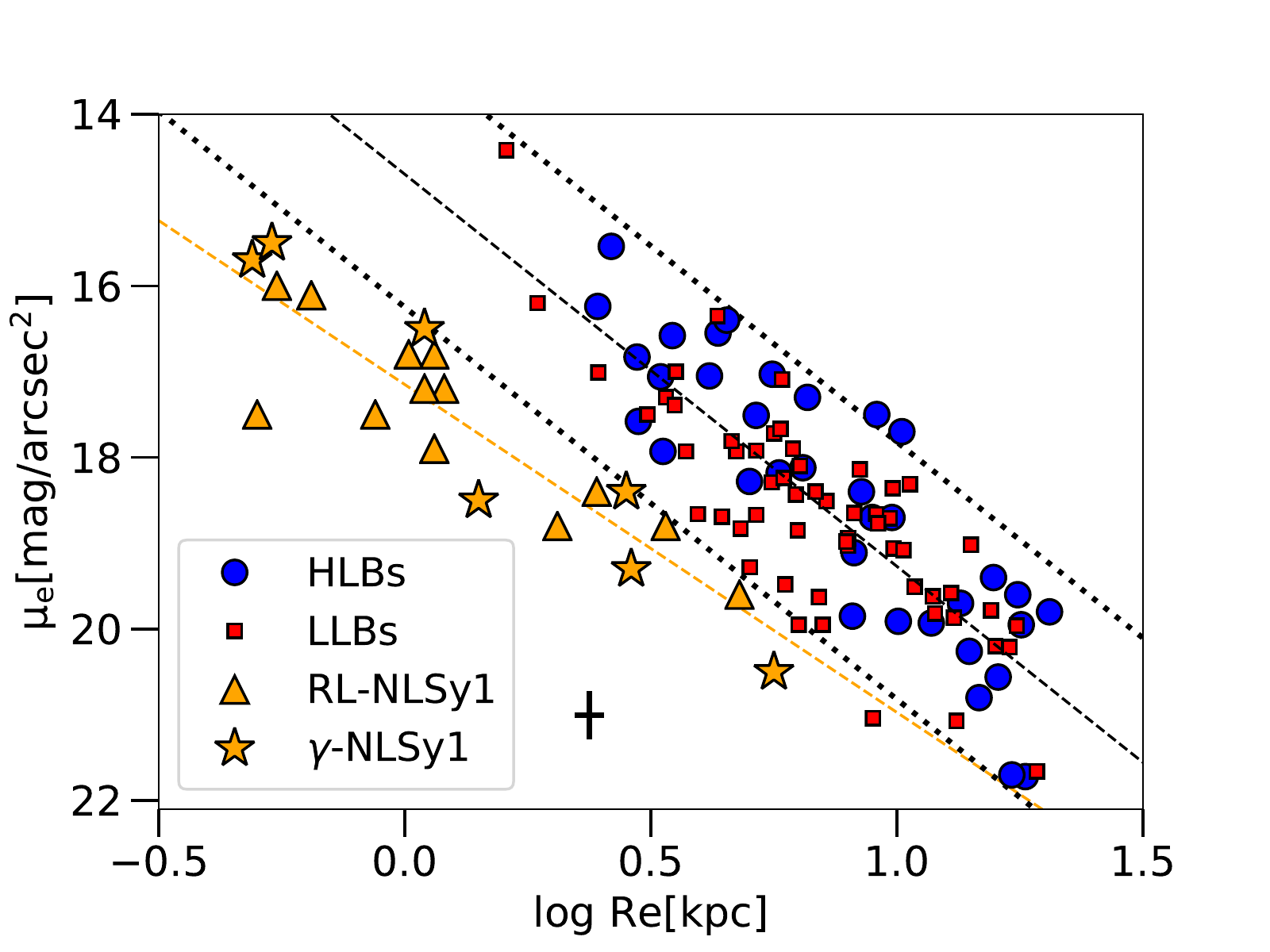}
 \caption{The Kormendy relation in K--band. Symbols are explained in the figure. 
 The bulge effective radius ($\mathrm{log R_e}$) is plotted versus the surface brightness at that
 radius ($\mathrm{\mu_e}$). A typical error bar is shown in the lower left corner. Overlayed are a sample of high--luminosity blazars (FSRQs) and low--luminosity blazars (BL Lacs objects) from \citet{olguin_2016}, all of them hosted by giant elliptical galaxies. For the blazars sample, we show the 95 per cent prediction bands (dotted lines) and the best linear fit (black dashed line) relation. 
 %[$\mu_e=(4.6 \pm 0.3)log R_e + (14.7 \pm0.3)]$. 
 For the NLSy1s (the 20 detected galaxies in our sample with K--band observations. Missing are 2 with unreliable fittings, 6 undetected galaxies and 1 without K--band observations), we show the best linear fit (yellow dashed line). There are no NLSy1s inside the 
 95 per cent prediction bands of the Kormendy relation, suggesting that they are hosted by disc--like bulges. A cosmological surface brightness dimming correction of the form $\mathrm{(1+z)^{-4}}$ was applied to all targets (in this work and in \citealt{olguin_2016})}
 \label{fig:kormendy2}
 \end{figure}

 In figure \ref{fig:kormendy2}, we explore the Kormendy relation by plotting the effective 
 radius ($\mathrm{R_e}$) versus the surface brightness at the bulge effective radius ($\mathrm{\mu_e}$) of 
 the detected hosts with K--band observations \footnote{In contrast to this work, the original 
 Kormendy relation, uses the average surface brightness within the bulge effective radius.}. 
 Together, we plot the results of the blazars hosts analysis from \citet{olguin_2016}. 
 The blazars hosts show a statistically significant correlation between $\mathrm{R_e}$ and $\mathrm{\mu_e}$ 
 (the Kormendy relation). On the other hand, the NLSy1s in the sample show a shallower trend. 
 The $\mathrm{95\%}$ prediction bands of the correlation are 
 represented by dotted lines. Since the aim of the prediction bands is to encompass the 
 likely values of future observations from the same sampled population, we might say that 
 it is most likely that the hosts that lie below the lower prediction band are not classical bulges. We 
 can see that all (20 with host detections, K--band observations and inside the 
 'correctly retrieved' area) RL--NLSy1s  lie below the Kormendy relation and below the 95\% prediction bands.
 If errors are taken into account, 4 NLSy1s might be consistent with classical bulges and 16 
 are certainly disc--like bulges.  
 
 Although these results suggest disc--like systems as RL--NLSy1s hosts, 
 some RL--NLSy1s might be hosted by classical--bulges. Therefore, further 
 studies of their stellar populations and kinematics, using integral 
 field spectroscopy (IFS), will help in understanding their
 nature. This result is not surprising, since the prevalence of disc--like bulges in this 
 type of AGN hosts have been previously found (e.g. \citealt{orban_2011,marthur_2012}). However, very little is known about the presence 
 of radio jets launched from disc galaxies galaxies. In the past, only a small number of disc--like 
 systems were found to be radio galaxies 
 (e.g. \citealt{ledlow_1998,hota_2011,Morganti_2011,Bagchi_2014,kaviraj_2015,mao_2015,singh_2015}). However, only 
 three (included in this work) have been found to, additionally, be $\gamma-$rays emitters 
 \citep[][1H0323+342,PKS2004-447, and FBQSJ1644+2619, respectively]{leontavares_2014,kotilainen_2016,olguin_2017}.\smallskip

 \begin{table*}
\centering
 \begin{minipage}{175mm}
 \caption{Parameters of the host galaxies of the NLSy1 sample derived from GALFIT 2D analysis. All K--corrections were performed using the 
\href{http://kcor.sai.msu.ru/}{K-corrections calculator} \citep{chilingarian_2010, chilingarian_2012}}
\label{table:galfit}
\resizebox{1.0\textwidth}{!}{%
  \begin{tabular}{@{}lccccccccccccc@{}}
  \hline
  Name	&	Band	&	$M_{nuc}$	&	$M_{bulge}$	&	$M_{disc}$ & Re (kpc)	&	$\mu_e$	&	$n$	&	$r_{s}$ & $f_{bar}$& $r_{bar}$(kpc) & $\chi_{best}$	&	$\chi_{best}/\chi_{psf}$	& Int\\
  (1)	&	(2)	&	(3)	&	(4)	&	(5)	&	(6)	&	(7)	&	(8)	&	(9)	& (10)& (11)& (12)& (13)& (14)\\
  \hline
  \multirow{2}{*}{1H0323+342$^\gamma$}	&	\multirow{2}{*}{J/K}	&	$-$23.00/$-$24.74	&	$-$22.53/$-$24.15	&	$-$23.49/$-$24.93 &1.48/1.40	&	16.89/15.14	&	0.88/1.24	&3.36/3.15 &\multirow{2}{*}{--} &	\multirow{2}{*}{--}& 1.280	&	\multirow{2}{*}{--} & \multirow{2}{*}{Y}	\\*
  	&		&	(0.01/0.01)	&	(0.01/0.01)	&	(0.01/0.01)	&	(0.01/0.01)	&	(0.01/0.01)	&	(0.002)	&	\\
  	
   \multirow{2}{*}{SBS0846+513$^\gamma$}	&\multirow{2}{*}{J/K}	&	$-$27.14/$-$28.99	&	\multirow{2}{*}{$>-23.89/> -25.68$}	& \multirow{2}{*}{--} &	\multirow{2}{*}{--} 	&	\multirow{2}{*}{--}	&	\multirow{2}{*}{--}	&\multirow{2}{*}{--} 	&\multirow{2}{*}{--} 	& \multirow{2}{*}{--}& 1.646/2.296	& \multirow{2}{*}{1}& \multirow{2}{*}{Y}	\\*
   	&		&	(0.36/0.38)	&		&		&		&	&	&	&	&	&(0.001/0.002) &\\

   \multirow{2}{*}{J093241+53063$^\gamma$}	&	\multirow{2}{*}{K}	&	$-26.90$	&	$-25.25$	& \multirow{2}{*}{--}&	5.57	&	20.50	&	1.21 &	\multirow{2}{*}{--}&	\multirow{2}{*}{--}&\multirow{2}{*}{--}&1.167	&	\multirow{2}{*}{0.40}	&\multirow{2}{*}{Y}\\*
   	&		&	(0.30)	&	(0.39)	&&	(0.58)	&	(0.48)	&	(0.70)	&	&&&(0.001)	&		\\

   \multirow{2}{*}{PMNJ0948+002$^\gamma$$^\times$}	&	\multirow{2}{*}{J/K}	&	$-27.39/-29.50$	&	$-24.53/-26.91$	&$-25.19/-27.59$ &	1.82/0.90	&	19.5/17.5	&	1.00/1.50	& 10.04/18.66& \multirow{2}{*}{--}&	\multirow{2}{*}{--}&1.718	&	\multirow{2}{*}{0.36}	&\multirow{2}{*}{N}\\*
   	&		&	(0.40/0.39)	&	(0.39/0.41)	&(0.40/0.41) &	(0.33/0.36)	&	(0.33/0.30)	&	(0.28/0.28)	&(0.36/0.38)	&&&(0.002)	&		\\

   \multirow{2}{*}{J095820+322401$^\gamma$$^\times$}	&	\multirow{2}{*}{K}	&	$-24.11$	&	$-22.91$	& $-25.94$&	1.47	&	16.76	&	0.60	&	10.04&\multirow{2}{*}{--}&\multirow{2}{*}{--}&1.200	&	\multirow{2}{*}{0.35} &\multirow{2}{*}{Y}\\*
   	&		&	(0.30)	&	(0.36)	&(0.48)&	(0.41)	&	(0.50)	&	(0.89)	&	(0.36)&&&(0.001)	&		\\
   	
   \multirow{2}{*}{FBQS J1102+2239}	&	\multirow{2}{*}{J/K}	&	$-25.09/-26.78$	&	$-23.78/-25.41$	& $-23.53/-25.74$ &	4.08/4.78	&	21.5/19.6	&	1.00/1.00	&11.06/16.24	&\multirow{2}{*}{--}&\multirow{2}{*}{--}&1.200	&	\multirow{2}{*}{0.32} &\multirow{2}{*}{Y}	\\*
   	&		&	(0.28/0.29)	&	(0.32/0.30)	&	(0.31/0.34)&(0.42/0.40)	&	(0.33/0.31)	&	(0.31/0.31)	&(0.41/0.40)&&&	(0.002)	&		\\
   		
   \multirow{2}{*}{IRAS11598-0112}	&	\multirow{2}{*}{J/K}	&	$-23.43/-25.69$	&	$-21.79/-24.66$	&$-23.30/-24.98$&	0.79/0.87	&	21.5/16.5	&	1.12/1.08	&2.24/2.47 &\multirow{2}{*}{0.12/0.09}&\multirow{2}{*}{10.46/10.46}&1.192/1.467	&	\multirow{2}{*}{0.30/0.28}	&\multirow{2}{*}{Y}\\*
   	&		&	(0.39/0.37)	&	(0.38/0.37)	&	(0.39/0.38)&(0.30/0.31)	&	(0.32/0.31)	&	(0.31/0.30)	&	0.50/0.48 &&&(0.003/0.003)	&		\\
   	
   \multirow{2}{*}{RXSJ12002-0046}	&	\multirow{2}{*}{J/K}	&	$-22.51/-24.01$	&	$-22.79/-24.22$	&	$-22.99/-24.16$ &1.39/1.16	&	18.5/16.8	&	1.10/1.08&4.50/4.15	&\multirow{2}{*}{--}&\multirow{2}{*}{--}&1.170/191	&	\multirow{2}{*}{0.38/0.38}	&\multirow{2}{*}{Y}\\*
   	&		&	(0.35/0.37)	&	(0.38/0.36)	&(0.38/0.38)	&(0.30/0.31)	&	(0.30/0.30)	&	(0.30/0.30)	&0.30/0.31&&&(0.002/0.002)	&\\
   \multirow{2}{*}{J12176+6546}	&	\multirow{2}{*}{J/K}	&	$-24.75/-26.86$	&	\multirow{2}{*}{$> -23.56/> -24.65$}	&\multirow{2}{*}{--}&	\multirow{2}{*}{--} 	&	\multirow{2}{*}{--}	&	\multirow{2}{*}{--}	&\multirow{2}{*}{--}	&\multirow{2}{*}{--}&\multirow{2}{*}{--}&1.241/1.558	&	\multirow{2}{*}{1.00/1.00} &\multirow{2}{*}{N}	\\*
   	&		&	(0.37/0.38)	&		&		&		&		&	&&&&(0.002/0.002)	&		\\
  
   \multirow{2}{*}{4C+04.42$^\gamma$}	&	\multirow{2}{*}{J}	&	-27.93	&	\multirow{2}{*}{$> -25.88$}	&	\multirow{2}{*}{--} 	&	\multirow{2}{*}{--}	&	\multirow{2}{*}{--}	&\multirow{2}{*}{--}&\multirow{2}{*}{--}&\multirow{2}{*}{--}&\multirow{2}{*}{--}&	1.200	&	\multirow{2}{*}{1.00}	&\multirow{2}{*}{N}\\
   	&		&	(0.37)	&	\multirow{2}{*}{--}	&	&	&		&&& &&	(0.002)	&		\\

   \multirow{2}{*}{RXSJ12278+3215}	&	\multirow{2}{*}{J/K}	&	$-$24.44/-26.24	&	\multirow{2}{*}{$> -23.10 / > -25.94$}	&	\multirow{2}{*}{--} 	&	\multirow{2}{*}{--}	&	\multirow{2}{*}{--}	&\multirow{2}{*}{--}&\multirow{2}{*}{--}&\multirow{2}{*}{--}&\multirow{2}{*}{--}&	2.030	&	\multirow{2}{*}{1.00}&\multirow{2}{*}{N}	\\*
   	&		&	(0.38/0.38)	&&		&		&		&&&&&	(0.001)	&		\\

   \multirow{2}{*}{SDSS J124634.65+023809$^\gamma$}	&	\multirow{2}{*}{K}	&	$-$25.01	&	$-$24.44	& -25.02&	2.85 	&	18.4	&	0.81	&	6.28  &	\multirow{2}{*}{--}&	\multirow{2}{*}{--}&1.188	&	\multirow{2}{*}{0.36}	&	\multirow{2}{*}{Y}\\*
   	&		&	(0.41)	&	(0.48)	&	(0.39)&(0.73)	&	(0.78)	&	(0.70)	&	(0.30)&&&(0.001)	&		\\
   \multirow{2}{*}{J13374+6005}	&	\multirow{2}{*}{J/K}	&	$-22.84/-24.49$	&	$-21.88/-23.54$	&$-22.30/-23.15$	&1.11/1.15	&	19.5/17.9	&	1.10/1.00	&	2.42/3.16&\multirow{2}{*}{0.14/--} &\multirow{2}{*}{11.54/--}&1.167/1.205	&	\multirow{2}{*}{0.36}&\multirow{2}{*}{N}	\\*
   	&		&	(0.37/0.39)	&	(0.39/0.41)	&	(0.39/0.40)&(0.28/0.30)	&	(0.29/0.31)	&	(0.32/0.35)	&	(0.30/0.34)& &&(0.001/0.002)	&		\\
   \multirow{2}{*}{J14033+0222}	&	\multirow{2}{*}{J/K}	&	$-21.91/-24.18$	&	$-22.98/-24.38$	&-22.89/-24.60&	3.98/3.39	&	19.7/18.8	&	1.11/1.05	&	7.26/6.70&\multirow{2}{*}{0.27/0.27}&\multirow{2}{*}{13.29/12.51}&1.168/1.172	&	\multirow{2}{*}{0.37}	&\multirow{2}{*}{N} \\*
   	&		&	(0.37/0.38)	&	(0.39/0.39)	&(0.41/0.42)&	(0.33/0.35)	&	(0.30/0.30)	&	(0.30/0.30)	&	0.37/0.35&&&(0.001/0.001)	&		\\
  
   \multirow{2}{*}{J142106+385522$^\gamma$}	&	\multirow{2}{*}{K}	&	$-26.63$	&	$-25.68$	& $-25.94$ &	2.89	&	19.3	&	1.11 	&25.08	&\multirow{2}{*}{--} &\multirow{2}{*}{--} &1.185	&	\multirow{2}{*}{0.31} &\multirow{2}{*}{Y}	\\*
   	&		&	(0.35)	&	(0.37)	&	(0.41)&(0.46)	&	(0.42)	&	(0.38)	&(0.57)	&& &(0.001)	&		\\*

   \multirow{2}{*}{J14506+5919}	&	\multirow{2}{*}{J/K}	&	$-22.45/-24.77$	&	$-21.90/-24.14$	&$-22.80/-25.01$ &	0.74/0.56	&	20.5/16.00	&	1.10/1.15	&4.40/4.84	&\multirow{2}{*}{0.22/0.22}&\multirow{2}{*}{7.98/9.14}&1.161/1.175	&	\multirow{2}{*}{0.34/0.32}	&\multirow{2}{*}{Y}\\*
   	&		&	(0.36/0.39)	&	(0.39/0.40)	&	(0.39/0.40)	&(0.37/0.39)&	(0.31/0.30)	&	(0.30/0.39)	&	(0.39/0.39)&&&(0.001/0.001)	&		\\   
   \multirow{2}{*}{B31441+476$^\gamma$}	&	\multirow{2}{*}{K}	&	-28.48	&		\multirow{2}{*}{$>-27.66$}	&	\multirow{2}{*}{--} 	&	\multirow{2}{*}{--}	&	\multirow{2}{*}{--}	&\multirow{2}{*}{--}&\multirow{2}{*}{--}&\multirow{2}{*}{--}&\multirow{2}{*}{--}&1.276	&	\multirow{2}{*}{1.00}&	\multirow{2}{*}{N}	\\*
   	&		&	(0.40)	&		&		&		&		&	&&&&(0.002)	&		\\ 	
   \multirow{2}{*}{PKS1502+036$^\gamma$$^*$}	&	\multirow{2}{*}{J/K}	&	$-24.17/-25.12$	&	$-23.80/-25.96$	& $-24.28/-26.30$	&1.05/0.49	&	15.5/15.7	&	1.06/1.15	&	4.11/4.17&\multirow{2}{*}{--}	&\multirow{2}{*}{--}	&1.577/1.236	&	\multirow{2}{*}{0.35/0.34}	& \multirow{2}{*}{Y}\\*
   	&		&	(0.38/0.39)	&	(0.39/0.38)	&	(0.39/0.41) &(0.64/0.70)	&	(0.36/0.39)	&	(0.60/76)	& (0.30/0.30)&	&&(0.001/0.001)	&		\\
   \multirow{2}{*}{SBS 1517+520}	&	\multirow{2}{*}{J/K}	&	$-24.51/-27.42$	&	$-24.33/-26.54$	&	-24.04/-26.32 &2.44/2.44	&	19.1/18.4	&	1.05/1.10	&8.66/5.75&\multirow{2}{*}{0.09/0.09}	& \multirow{2}{*}{10.26/10.26}& 1.201/1.241	&	\multirow{2}{*}{0.35/0.34} &	\multirow{2}{*}{Y}	\\*
&		&	(0.39/0.38)	&	(0.38/0.39)	&	(0.37/0.37)&(0.32/0.33)	&	(0.32/0.32)	&	(0.32/0.32)	&(0.30/0.33)	&&&(0.002/0.002)	&		\\   	
   \multirow{2}{*}{B2 1546+35A}	&	\multirow{2}{*}{J/K}	&	$-24.78/-26.38$	&	$-24.19/-24.74$	& $-23.12/-25.69$ &	0.96/2.06	&	16.5/18.8	&	1.18/0.98	&3.99/10.52	&\multirow{2}{*}{--} &\multirow{2}{*}{--}&1.161/1.153	&	\multirow{2}{*}{0.32}&\multirow{2}{*}{Y}	\\*   
   	&		&	(0.32/0.30)	&	(0.29/0.32)	&	(0.31/0.31) &(0.84/0.33)	&	(0.90/0.32)	&	(0.62/0.41)	&	(0.41/0.41)&&&(0.001/0.001)	&		\\   		
 
   \multirow{2}{*}{J16290+4007}	&	\multirow{2}{*}{J/K}	&	$-23.85/-25.86$	&	\multirow{2}{*}{$> -22.64 / -24.35$}	&	\multirow{2}{*}{--} 	&	
   \multirow{2}{*}{--}	&	\multirow{2}{*}{--}	&\multirow{2}{*}{--} &\multirow{2}{*}{--} &\multirow{2}{*}{--} &\multirow{2}{*}{--} &	1.118	&	\multirow{2}{*}{1.00}&\multirow{2}{*}{Y}	\\*
   
   	&		&	(0.37/0.39)	&		&		&		&		&&&&&	(0.002)	&	&	\\
   	
   \multirow{2}{*}{RXSJ16333+4718}	&	\multirow{2}{*}{J/K}	&	$-22.90/-24.40$	&	$-22.35/-23.22$	&$-23.01/-24.15$	& 3.61/0.50	&	22.5/17.5	&	1.10/1.21	&	8.50/2.83 &\multirow{2}{*}{0.14/0.12}&\multirow{2}{*}{5.04/5.04}&1.299/1.310	&	\multirow{2}{*}{0.32}	&\multirow{2}{*}{Y}\\*
   	&		&	(0.37/0.37)	&	(0.39/0.38)	&	(0.39/0.39)&(0.31/1.58)	&	(0.32/1.40)	&	(0.35/0.55)	&	0.34/0.34&&&(0.001/0.001)	&		\\

   \multirow{2}{*}{2E1640+5345}	&	\multirow{2}{*}{J/K}	&	$-21.69/-23.73$	&	$-23.40/-24.52$	&	$-24.21/-25.14$&0.45/0.64	&	17.5/16.6	&	1.10/1.09	&	4.50/5.72& \multirow{2}{*}{0.07/0.08} & \multirow{2}{*}{7.65/6.17}   &1.178/1.133	&	\multirow{2}{*}{0.35}  & \multirow{2}{*}{N}	\\*
   	&		&	(0.40/0.39)	&	(0.37/0.37)	&(0.39/0.39)	&(0.32/0.32)	&	(0.40/0.38)	&	(0.32/0.32)	&	(0.31/0.31)&&&(0.001/0.001)	&		\\

   \multirow{2}{*}{J16410+3454}	&	\multirow{2}{*}{J/K}	&	$-21.27/-24.35$	&	$-23.55/-24.65$	& $-24.18/-25.55$ &	0.83/0.98	&	17.5/16.8	&	2.24/1.53	&	4.34/4.01 &\multirow{2}{*}{0.22/0.22}&\multirow{2}{*}{21.20/21.10}&1.192/1.169	&	\multirow{2}{*}{0.30} &\multirow{2}{*}{Y}	\\*
   	&		&	(0.39/0.38)	&	(0.39/0.39)	& (0.38/0.39)&	(0.49/0.46)	&	(0.34/0.38)	&	(0.43/0.42)	&	(0.35/0.33)&&&(0.001/0.001)	&		\\

   \multirow{2}{*}{FBQSJ1644+2619$^\gamma$}	&	\multirow{2}{*}{J/K}	&	$-22.51/-24.80$	&	$-21.41/-24.21$	&$-22.90/-24.85$	&0.96/1.10	&	17.5/16.5	&	1.80/1.90	&	6.65/7.68& \multirow{2}{*}{-/0.17}& \multirow{2}{*}{-/8.13}&1.160	&	\multirow{2}{*}{0.37}&\multirow{2}{*}{Y}	\\*
   	&		&	(0.43/0.24)	&	(0.40/0.32)	&	(0.35/0.38)&(0.32/0.34)	&	(0.30/0.31)	&	(1.31/1.35)	&(0.45/1.14)	 & & &(0.027)	&		\\

   \multirow{2}{*}{B31702+457}	&	\multirow{2}{*}{K}	&	$-23.88$	&	$-25.09$	&	$-23.63$ & 1.21	&	17.2	&	1.03	&	5.97& \multirow{2}{*}{--}&\multirow{2}{*}{--}&2.021	&	\multirow{2}{*}{0.35}	& \multirow{2}{*}{N}	\\*
   	&		&	(0.28)	&	(0.32)	&	(0.30)&(0.36)	&	(0.43)	&	(0.41)	&(0.29)&&&(0.003)	&		\\  	
   	
   \multirow{2}{*}{J17221+5654}	&	\multirow{2}{*}{J/K}	&	$-23.84/-25.64$	&	$-23.95/-24.95$	& $-23.12/-24.35$ &	0.71/1.11	&	18.3/17.2	&	2.10/2.12	&	4.34/4.01& \multirow{2}{*}{--}&\multirow{2}{*}{--}&1.164/1.172	&	\multirow{2}{*}{0.27/0.30} &	\multirow{2}{*}{Y}	\\
   	&		&	(0.35/0.36)	&	(0.38/0.39)	&	(0.39/0.39) &(0.34/0.36)	&	(0.40/0.35)	&	(0.33/0.32)	&	(0.33/0.32)&&&(0.001/0.001)	&		\\*

   \multirow{2}{*}{PKS2004$-$447$^\gamma$$^*$}	&	\multirow{2}{*}{J/K}	&	$-22.56/-24.79$	&	$-22.91/-24.50$	&$-23.05/-24.40$	&0.72/0.53	&	16.6/15.5	&	1.15/1.08	&	3.65/2.51& \multirow{2}{*}{-/0.13} &\multirow{2}{*}{-/7.59}&1.403	&	\multirow{2}{*}{0.29}&\multirow{2}{*}{Y}	\\
   	&		&	(0.32/0.25)	&	(0.29/0.28)	&(0.27/0.30)	&(0.34/0.42)	&	(0.20/0.19)	&	(0.10/0.35)	&	(0.32/0.32)&&&(0.025)	&		\\
   	
   \multirow{2}{*}{IRAS22453$-$1744}	&	\multirow{2}{*}{J}	&	$-23.34$	&	$-24.10$	&\multirow{2}{*}{--}&	2.28	&	17.5	&	0.90	&\multirow{2}{*}{--}&\multirow{2}{*}{--}&\multirow{2}{*}{--}&	1.176	&	\multirow{2}{*}{0.27}&\multirow{2}{*}{Y}	\\
   	&		&	(0.35)	&	(0.36)	&&	(0.29)	&	(0.35)	&	(0.30)	&&&&(0.001)	&		\\*	 	
\hline
\end{tabular}}\\
\scriptsize {Column (1) gives the galaxy name; 
(2) the observed band; 
(3), (4)  and (5) the nuclear, bulge and disc absolute magnitudes
for the best-fit model in the observed band. When the galaxy is not detected, 
we determine an upper limit by simulating a bulge with average parameters. 
(6) Bulge effective radius;
(7) the bulge model surface brightness at the effective radius; 
(8) the bulge model S\'ersic index; 
(9) the disc model scale length;
(10) bar strength;
(11) bar length;
(12) the reduced $\chi^{2}$ for the best-fit model; 
(13) the ratio between best-fit $\chi^{2}$ and the PSF-fit $\chi^{2}$; 
(14) if the galaxy shows some sign of interaction (Y), otherwise (N).\\
$^{*}$ \scriptsize{Galaxies observed using the ISAAC on the ESO/VLT}\\ 
$^\gamma$ \scriptsize{$\gamma-$ray emitting NLSy1 galaxies}\\
$^\times$ \scriptsize{Galaxies with unreliable fittings according to our simulations (see Section \ref{section:simulations}).}}
\label{table:galfit}
\end{minipage}
\end{table*}

 The uniqueness of the hosts of NLSy1s 
 (and nearby environment), might hint at how these galaxies acquired such 
 properties. 
 Spiral arms, bar incidence, interactions evidence, etc., could shed some light on the fueling 
 mechanisms 
 needed by the central supermassive black hole to form and develop powerful relativistic jets. 
 Therefore, in the following sections we discuss on the specific features that the hosts of the 
 galaxies in the sample show.\smallskip
 
 \subsection{Bar frequency} \label{section:bars}
 Through a simple visual inspection, some galaxies reveal the presence of a bar in 
 their brightness distributions. 
 However, in order to have a quantitative identification of these bars, we adopt an analysis based on the ellipse 
 fit to the galaxy isophotes, where radial variations of ellipticity ($\epsilon$) and position angle (P.A.), 
 exhibit the existence, ellipticity and extent of the bar. Moreover, some bars detected by this way, might 
 not be obvious at a glance. The detection of bars using this method consists on finding a local maximum 
 in $\epsilon$, which indicates the bar end. Along the bar, the P.A. should remain constant, thus along 
 the suspected bar, the P.A. should not change much (typically $\mathrm{\Delta \textrm{PA}\lesssim 20^\circ}$, \citealt{wozniak_1995, 
 jogee_1999,mendendez_2007}). At larger radius, further away from the bar end, we should measure the ellipticity and 
 P.A. of the disc then, the ellipticity should drop (at least 0.1; $\mathrm{\Delta\epsilon \ge 0.1}$) and most 
 likely the P.A. will change. In Appendix \ref{appendix_B} 
 we show the plots of P.A. and ellipticity versus radius of 10 of the galaxies that fulfill this criteria
 \citep[the other two galaxies in the sample with detected bars are shown in][]{kotilainen_2016, olguin_2017}. 
 Thus, the detected hosts with bars represent 52\% (77\% RL--NLSy1s and 20\% $\gamma-$NLSy1s) 
 of the galaxies in the sample, in comparison, \cite{laurikainen_2004} finds bars in 62\%$\pm9\%$ of Seyfert galaxies. 
 It must be stressed that, although near-IR imaging (specially K--band) provides a reliable 
 assessment of the bar fraction, we might miss some bars given the relative high redshift of some of 
 the galaxies in the sample (e.g. J17221+5654 is the galaxy with the highest redshift $\mathrm{z=0.426}$ 
 and a bar detection).\smallskip
 
 In addition to finding bars, the radial variations of ellipticity and P.A., help in estimating 
 their ellipticity and length. Ellipticity has been shown to be a good bar strength indicator 
 \citep{laurikainen_2002}. \citet{abraham_merrifield_2000} defined a bar strength parameter given by 
 
 \begin{equation}
 \mathrm{f_{bar}=\frac{2}{\pi}\left(arctan\left(1-\epsilon_{bar}\right)^{-1/2}-arctan\left(1-\epsilon_{bar}\right)^{1/2}\right)},
 \label{eqn:bar_strength}
 \end{equation}
 
 where $\mathrm{\epsilon_{bar}}$ is the bar ellipticity at the bar end. Using this parameter, we find that, for 
 our sample, $\mathrm{f_{bar}=0.13\pm0.06}$. By comparing this value with the findings of 
 \citealt{laurikainen_2007} (average $\mathrm{f_{bar}=0.20\pm0.03}$, from 216 galaxies observed in the NIR, 
 including all Hubble types), 
 we note that the bar strengths for our NLSy1s sample is rather low. By contrast, we find that the
 average bar length of our sample ($\mathrm{r_{bar}=9.8kpc\pm1.8}$) is large when compared either with late 
 type galaxies or early type galaxies \citep[$\mathrm{\sim 0.5-3.5kpc}$ and $\mathrm{\sim1-10kpc}$, respectively; ][]{erwin_2005}.
We note that, even if we assume that the galaxies with no bars detected in our sample have the shortest bars (0.5 kpc), the average bar length would be 5.9 kpc, still larger than the average for late type galaxies (2 kpc), early type galaxies (5.5 kpc) or both early 
 and late type galaxies (3.75 kpc). These results might hint to the necessary conditions to properly channel the fuel towards the center 
 of the galaxy to feed the black hole and trigger its activity. According to bar models by 
 \citet{athanassoula_1992}, weak bars promote the inflow of gas toward the inner Lindblad resonance (ILR), 
 and forms a nuclear ring. As long as the bar pattern speed remains low, the ILR is kept, which occurs 
 provided that bars are long. If dynamical instabilities via gradual build-up of material show, 
 material from that nuclear ring would flow inward and trigger the black hole activity, as hypothesize by
 \citet[][]{laurikainen_2002}. Low and steady evolution (secular evolution) might, thus, 
 be capable of producing AGN as powerful as to launch radio jets.\smallskip
 
 \subsection{Mergers and galaxy interactions}
 
 Secular evolution driven by stellar asymmetries (i.e. bars, lopsidedness, spiral patterns 
 and other coherent structures) can be strengthened by external processes such as tidal 
 interactions and mergers \citep{mapelli_2008, reichard_2009}. In our sample, 62\% of the 
 galaxies show some sign of merger, interaction or off centered components (9 $\gamma-$NLSy1s 
 and 9 RL--NLSy1s, including both detected and not detected hosts). This result is important 
 since both, observations and simulations suggest that AGN activity is closely related to galaxy 
 interactions and mergers \citep[e.g.,][]{dimatteo_2003,ellison_2011,silverman_2011,koss_2012,ellison_2013,capelo_2015}.\smallskip
 
 Particularly, in the case of the $\gamma-$NLSy1s, we note that only 3 out of 12 $\gamma-$NLSy1s do 
 not show signs of interactions (PMNJ0948+0022, $z=0585$; 4C+04.42, $z=0.996$ and B3 1441+476, $z=705$), 
 considering their high redshift and that only one of these hosts is detected. The large fraction 
 of interactions in the $\gamma-$NLSy1s of the sample (75\% against 53\%, when compared with RL--NLSy1s), 
 largely consist of the host itself and another, significantly smaller, galaxy or faint tidal 
 feature (i.e. minor mergers; however, spectroscopic data is required to confirm the idea of these features as 
 interactions).\smallskip
 
 This result is important since simulations \citep[e.g.][]{qu_2011} show that 
 the angular momentum decreases more significantly when the stellar disc undergoes a minor merger than when 
 it evolves in isolation. Hence, the difference in power between RL--NLSy1s and $\gamma-$NLSy1s,
 might thus be the result of the difference between the processes that drive their evolution.
 In this way, our findings suggest secular evolution as a process capable of
 producing, not only radio, but also $\gamma-$ray emitting jets.\smallskip
 
Another important finding in our study is that, not only discs might be off-centered with respect to
the nucleus, also some bulges might . This behavior has not only been predicted by simulations \citep{hopkins_2012} but also previously observed. The first offset AGN reported was the Seyfert 1 galaxy NGC 3227 \citep{ mediavilla_1993}, where the region of broad emission lines is offset from the kinematic center by  $\sim 0.250$ kpc. Another important example is the low--luminosity AGN
NGC 3115 \citep{menezes_2014}, with an AGN located at a projected distance of $\sim 0.014$ kpc from the stellar bulge. Similarly, using GALFIT and observations from Chandra/ACIS and the Hubble Space Telescope, \cite{comerford_2015} analyzed a sample of 12 dual AGN candidates at $\mathrm{z<0.34}$  and discovered 6 systems that are either dual or offset AGNs with separations $\mathrm{\Delta x<10}$kpc. Finally, here we find a total of 6 systems (see Table \ref{separation}) where the stellar bulge is offset from the AGN by projected distances  $\mathrm{\Delta x<1.5}$kpc

 \begin{table}
\begin{center}
\caption{Summary of galaxies with offset AGNs.}\label{separation}
 \begin{tabular}{lccc}
 \hline
Galaxy	&	Redshift	&$\mathrm{\Delta x}$(arcsec)	&$\mathrm{\Delta x}$(kpc)\\
 \hline
RXSJ12002--0046 &0.179 &0.24 &0.73\\
J142106+385522 $^\gamma$ & 0.490 & 0.25&1.50\\
J14506+5919  &0.202& 0.23&0.76\\
B2 1546+35A & 0.479 &0.24&1.42\\
J17221+5654 & 0.426 &0.23 &1.30\\
IRAS 22453--1744&0.117 &0.71&1.50\\
 \hline
   \end{tabular}
\end{center}
\scriptsize {Column (1) gives the galaxy name; 
(2) the redshift of the system;
(3) and (4) the projected separation between the stellar bulge and the AGN in arcsec and kpc, respectively. The typical error is $\pm 0.16''$, derived as described in section \ref{errors}.\\ 
$^\gamma$ \scriptsize{$\gamma-$ray emitting NLSy1 galaxy}\\
}
 \end{table}

This finding strongly suggests an important connection between AGNs and galaxy mergers. Two likely scenarios where the AGN is off--centered with respect to the stellar bulge are explained on the basis of  galaxy mergers. On the one hand, the black hole of the observed AGN and another (inactive) black hole form a binary system. The inactive black hole is located in the center of stellar bulge, and thus the AGN is offset with respect to it \citep{menezes_2014}. On the other hand, two black holes might have already coalesced which 
caused the formation of gravitational waves, which in turn, asymetrically push the system to shaped it to 
its current form \citep[][and references therein]{merritt_2006,blecha_2008,sundararajan_2010,blecha_2019}. 
This important finding might help in constraining SMBH-galaxy co-evolution theoretical studies and simulations 
where most of the times, a stationary central black hole is assume.
 
 \subsubsection{Notes on individual galaxies}
 Here, we provide a short description of the characteristics that each interacting 
 galaxy shows (for images, see Appendix \ref{appendix_A}).\smallskip
 
 \begin{itemize}
 
   \item 1H 0323+342. The closest $\gamma-$ray emitter NLSy1 galaxy. In this galaxy a ringed structure is seen which is interpreted as  ``suggestive evidence for a recent 
   violent dynamical interaction''  by \cite{leontavares_2014}, where an extensive discussion on this galaxy can be found.
   
   \item SBS 0846+513. The host galaxy of this $\gamma-$NLSy1 is not detected, therefore it was modeled 
   using a PSF. A bright and close companion ($\mathrm{<5}$arcsec) is clearly detected. Also, a spiral galaxy in the 
   foreground is observed.   
   
   \item J093241+53063. A source with a disc--like bulge, with S\'ersic index $n=1.21\pm0.70$. Although no disc is detected, a faint companion is. 
   
    \item J095820+322401. Although its relative high redshift ($\mathrm{z=0.53}$), we detect a faint close neighbor in this galaxy. When 
   analysed further, an even fainter disc is reveled in our analysis. The disc is off--centered with respect to the nucleus and bulge. 
   This, maybe due to the action exerted by its alleged neighbor. 
  
  \item FBQSJ1102+2239. This galaxy represents a classical encounter between two disc galaxies. Very similar both in J-- and K--bands, a remnant of the disc is detected in the 
   AGN host. The companion keeps an spiral arm and it is barely connected with the main AGN host. Another blob is observed, an HII region, which is part of the system 
   (according to optical spectra).
  
   \item IRAS 11598-0112. The AGN host is modeled using a disc--like bulge and a disc. However, two spiral-arm-like features are included in the model. More interestingly, another component 
 (probably a disky bulge) is detected inside the main disc.	
   
    \item SDSS J124634.65+023808. An AGN dominated galaxy modeled using a bulge and an exponential disc. It shows a close and faint feature which is fitted using an exponential disc.
  
   \item J14033+0222. An apparent simple barred galaxy. However, the disc is off-centered with respect to the AGN and bulge (in both, J-- and K--bands), which suggests to a not obvious interaction. 
   Since both, bar and disc have exponential profiles and the bar is faint, only one exponential was 
   needed to model both.
  
  \item J142106+385522. The host galaxy of this $\gamma-$NLSy1 was modeled using a bulge and a disc. In the image, a faint 
   tail--like feature (which was not modeled due to its intensity) is observed. Suggestion of an interaction are observed in its 
   off--centered disc. 
  
  \item PKS1502+036. A clear companion only visible in J--band. The companion seems close, 
  however, this galaxy does not show asymmetries as others in the sample. Although \citet{dammando_2018} 
  finds it hosted in an elliptical galaxy, we find a better fit using a disc--like host. 
   
   \item SBS1517+520. In this galaxy, the asymmetry of its surface brightness profile and its off centered disc (more 
   evident in the K--band), hint towards some type of disruption induced by an interaction.
   
   \item B2 1546+35A. When the host is decomposed into bulge and disc, the different parts are off centered with respect to each other. 
   While the host galaxy image, both in J-- and K--bands, looks similar, 
   when it is represented using its surface brightness profile, the two bands differ.
   
  \item RXSJ16333+4718. Two disc galaxies interacting. Seemingly, the companion is also face-on. Again, the host galaxy shows a greater effect of the interaction 
   on one of the bands. While in K--band, the bulge seems a bit off centered, in the J--band, both the disc and the bulge shows greater impact on its morphology. 
   
   \item J16410+3454. This galaxy shows a feature that emerges after the fitting of a bulge+disc+AGN model. This additional component is modeled using a S\'ersic profile with $\mathrm{n\approx1}$. The main disc seems off centered in the 
   radial profiles (maybe because of the effect of the interaction). 
   
   \item FBQS J1644+2619. In J--band, a ring feature shows, whose formation process might be that described by \cite{athanassoula_1997} 
   and that PKS 2004-447 might be undergoing. Besides, a faint disruption of the ring suggests an interaction. 
   On the other hand, in K--band, a bar is observed and the ring features is almost absent. An extensive discussion on these features is found on \cite{olguin_2016}.
  
  \item J17221+5654. The host galaxy was modeled using a bulge and a disc. In both J-- and K--bands a small companion is detected a few arcseconds away from the host center.
   The companion seems to be interacting with the main galaxy since the components of the AGN host are off centered; more evident in K--band.
  
   \item PKS 2004-447. This barred galaxy shows two faint spiral arms, one of which is more open. It is a very good example of part of the evolution of the simulations by 
   \cite{athanassoula_1997}, where the impact of a small companion on a barred galaxy leads to the formation of a ring. For a detailed 
   discussion on this galaxy see \citet{kotilainen_2016}. 
  
  \item IRAS 22453-1744. A bulge with a S\'ersic index $\mathrm{n=0.90\pm0.30}$ was used to model the host. However, the AGN and host are off-centered. The most likely reason is the close (merged) companion  with a complex morphology that is difficult to characterize.
  
   \end{itemize}

 \subsection{AGN, bulge and disc (J-K) colours}
 
 \begin{table}
 \centering
 \begin{tabular}{lccc}
 \\
 \hline
 Subsample	&	$(J-K)_{Bulge}$	&	$(J-K)_{Disc}$	&	$(J-K)_{Nuclear}$\\
 \hline
 All 	        	  &$1.7\pm0.7$ 	&$1.7\pm0.5$ 	& $2.0\pm0.5$ \\
 RL--NLSy1s 	    & $1.5\pm0.5$ & $1.6\pm0.6$ & $2.0\pm0.5$ \\
 $\gamma-$NLSy1s  &$2.1\pm0.5$ 	&$1.8\pm0.4$ 	& $1.9\pm0.5$\\
 \hline
   \end{tabular}
 \caption{Average $\mathrm{J-K}$ colours for the bulge, disc and nucleus of the host galaxies in the sample.}\label{table:colours}
 \end{table}

 Table \ref{table:colours} shows the average J-K colours of the main components of the host galaxies 
 in the sample (whenever they have both, J-- and K--band information).
 We see that the disc and nuclear J-K colours remain virtually unchanged whether the 
 subsample includes $\gamma-$NLSy1s or only RL--NLSy1s. We also note that, unexpectedly, 
 bulge and disc colors for RL--NLSy1s are the same within errors 
 \citep[i.e. bulges are as blue as discs, when disc are expected to be bluer,][]{moriondo_1998, seigar_1998}.
 This result is thus consistent with star--forming bulges, which imply large gas reservoirs.\smallskip
 
 On the other hand, the average J-K bulge colour changes depending
 on the subsample, being redder for $\gamma-$NLSy1s ($\mathrm{J-K=2.1\pm0.5}$). According to findings by 
 \cite{glass_1985,seigar_1998}, NIR colours $\mathrm{J-K\approx 2.0}$, could be the result of a dust-embedded AGN 
 or a nuclear starbursts (if the component is extended, i.e. in bulges). 
 Bulge reddening might be linked to the large fraction 
 of $\gamma-$NLSy1s showing signs of minor mergers. Thus, according to these results, interactions 
 are likely to play an important role in the nuclear activity of the galaxies in our sample.\smallskip
 
 %since high recent starburst activity could be reflected (by means of supergiants) in K--band.

 \begin{figure*}
 \centering
 \includegraphics[scale=0.45]{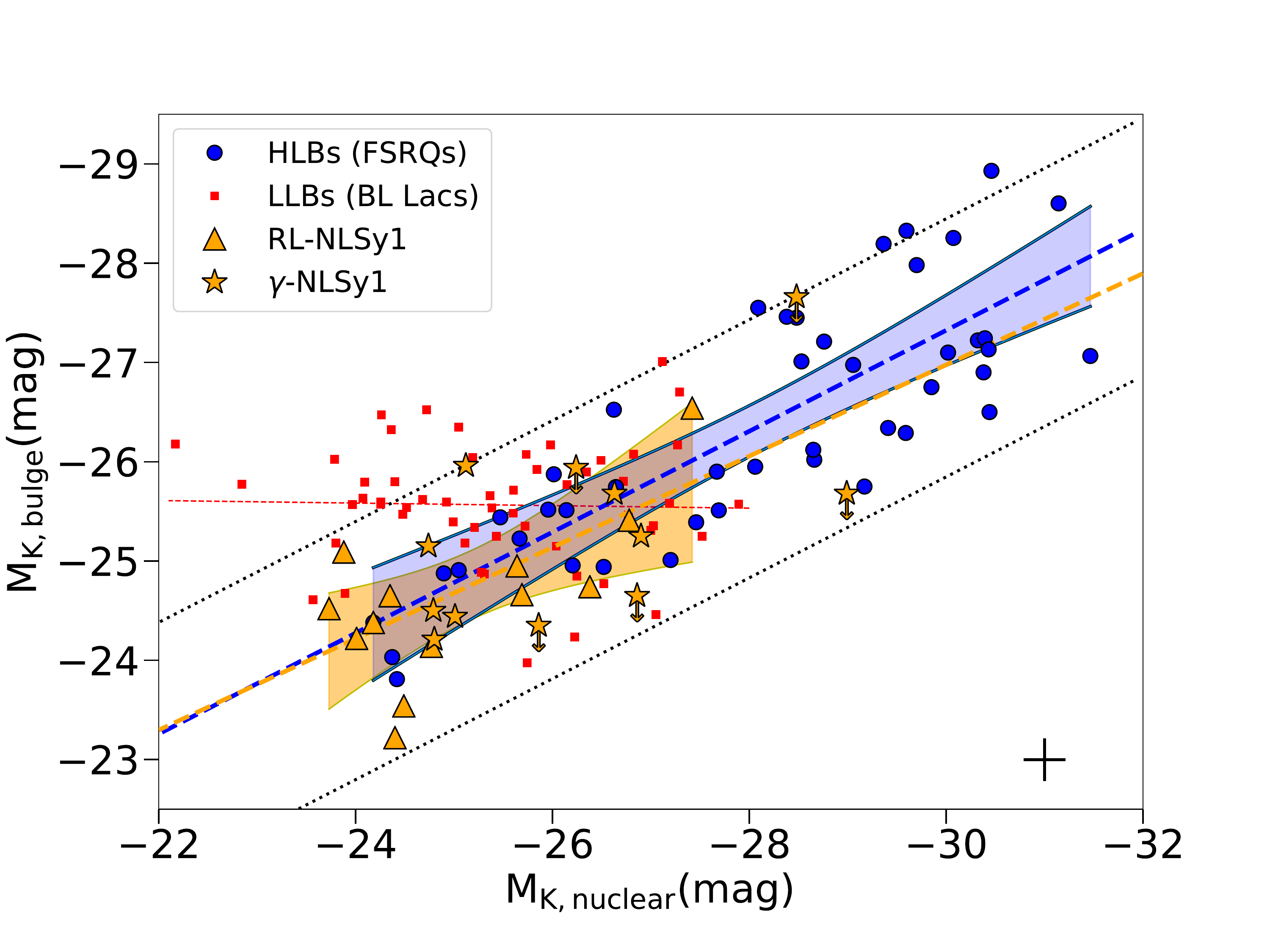}
 \caption{Plot of the nuclear K-band magnitude versus the bulge K-band magnitude for the host galaxies of the sample. 
 Symbols are explained in the figure. Upper limits for unresolved galaxies are shown as down arrows. We show the 25 detected and undetected 
 galaxies in our sample with K--band observations. Missing are 2 with unreliable fittings and 2 without K--band observations.
 Overlayed is a sample of blazars from \citet{olguin_2016}, where the galaxies were analysed in K--band. 
 For the blazars sample we show the best linear fits (dashed blue line for FSRQs and dashed red line for BL Lacs) and the 95 per cent prediction bands (dotted black lines). The 99 per cent confidence intervals are shown for FSRQs (blue shade) and NLSy1s (yellow shade).
 For the NLSy1s sample we show the best linear fit (green solid line). All K–corrections were
performed using the \href{http://kcor.sai.msu.ru/}{K-corrections calculator} \citep{chilingarian_2010, chilingarian_2012}
 }
 \label{fig:agn-host}
 \end{figure*}

 \section{AGN-host connection}
 In figure \ref{fig:agn-host}, we explore the $\mathrm{M_{K, nuclear}-M_{K, bulge}}$ plot for our sample. The first thing we
 notice is that most of the sample data points fall in the bottom left corner, where high luminosity blazars (HLB), i.e. 
 FSRQs and low luminosity blazars (LLB), i.e. BL Lacs coincide. 
 However, three $\gamma-$NLSy1s (only one with host galaxy detection) are brighter
 and lie $\sim3$mag apart from the group and two (also $\gamma-$NLSy1s) are fainter and lie $\sim1$mag 
 appart. At first glance the bulk of RL--NLSy1s, are part of either HLB or LLB. 
 A further analysis shows that, if we include the NLSy1s in the HLB sample, the best 
 linear fit marginally lies inside the 99\% confidence intervals of the FSRQs best linear fit. There is a statistically significant positive correlation for the FSRQs sample ($\mathrm{r=0.8}$, $\mathrm{p=4\times10^{-12}}$) which is kept virtually unchanged ($\mathrm{p\approx10^{-12}}$) when the 
 RL--NLSy1s are added to the FSRQs.\smallskip
 
 When we analyze the NLSy1s sample alone, a statistically significant 
 positive correlation is observed ($\mathrm{r=0.6}$, $\mathrm{p=1\times 10^{-3}}$).
 Whether the NLSy1s sample belongs to the FSRQs sample or not, its positive trend suggests 
 that their jets also stimulate starburst activity in their hosts.\smallskip
 
 The results presented above suggest either FSRQs and RL--NLSy1s accretion modes locate them 
 close to each other in the $\mathrm{M_{K, nuclear}-M_{K, bulge}}$ plot. Bearing in mind that both, 
 FSRQs and NLSy1s, are thought to acrete matter very efficiently via accretion discs, 
 this is expected.\smallskip 
 
 In order to rule out the scenario where RL--NLSy1s behave as LLB/BL Lacs 
 (since, independently of nuclear magnitude, BL Lacs show a narrow range of bulge magnitudes, 
 where the bulk of RL--NLSy1s bulge magnitudes also reside), we perform a two-dimensional Kolmogorov-Smirnov test for the bulge and nuclear magnitudes of BL Lacs and RL--NLSy1s. According to the results of this test ($\mathrm{p\approx3.0^{-4}}$) there is a statistically 
 significant difference between the samples, suggesting that, in the $\mathrm{M_{K, nuclear}-M_{K, bulge}}$ plot, 
 BL Lacs and RL--NLSy1s are not similar.
 \smallskip
 
 Previous studies, conducted to RL--NLSy1s, had already supported the idea that, when compared with blazars, 
 RL--NLSy1s are particularly similar to FSRQs \citep[e.g.][]{paliya_2013, foschini_2015, paliya_2016}. Provided that RL--NLsy1s are hosted by spiral galaxies, what our findings suggest, might be a substantive contribution since 
 blazars are known to be hosted by elliptical galaxies and powered by massive black holes.\smallskip
 
The positive trend in the NLSy1s and FSRQs samples clearly suggest a positive feedback scenario. The AGN outflows can induce star formation, both in the galactic disc through compression of molecular clouds \citet{silk_2013} or directly in the outflowing gas \citet{ishibashi_2012}.

 \section{Summary} \label{summary}
 We have presented near--infrared images of a sample of 29 radio--loud NLSy1 host galaxies, 
 12 of which are also classified as $\gamma-$NLSy1s. By thoroughly analyzing their 
 2D surface brightness distribution, we successfully detected 21 hosts (14 RL--NLSy1s and 7 $\gamma-$NLSy1s). 
 Our near--infrared study allowed us to compare the photometrical properties of RL--NLSy1s 
 with another type of AGN capable of launching powerful radio jets, namely blazars (both BL Lacs and FSRQs). 
 The main findings of our study are summarized below. 
 
 \begin{itemize}
  \item The photometrical properties derived by our 2D analysis for a sample of RL--NSLy1s suggests that,
  consistent with radio-quiet NLSy1s and opposite to the jet paradigm, powerful relativistic 
  jets can be launched from disc--like systems instead of elliptical galaxies or classical bulges.
  
  \item Secular evolution driven by the peculiar bar properties in our sample (long and weak) 
  might be responsible for channeling fuel towards the center of the galaxy to feed 
  the black hole and trigger the nuclear activity, whereas the nuclear activity 
  in $\gamma-$NLSy1s could be the result of a similar process enhanced by minor mergers.  
  
   \item RL--NLSy1s bulges and discs show the same average NIRcolour 
   $\mathrm{(J-K)_{Disc}=1.6\pm0.6}$ and $\mathrm{(J-K)_{bulge}=1.5\pm0.5}$. Since discs are expected to be bluer, 
   this result is consisting with star--forming bulges, suggesting large gas reservoirs.
   On the other hand, $\gamma-$NLSy1s bulges show an average NIR colour $\mathrm{(J-K)_{AGN}=2.1\pm0.5}$. 
   This reddening (with respect to RL--NLSy1s) suggests a nuclear starburts, probably, 
   linked to the large fraction of minor mergers shown by $\gamma-$NLSy1s, which in turn, 
   could make a difference between RL-- and $\gamma-$NLSy1s nuclear activity.
   
   \item We have discovered 6 systems showing an offset stellar bulge with respect to the AGN (with separations $\mathrm{\Delta<1.5}$kpc).
   This might be the result of a galaxy merger, strongly suggesting an important connection between AGNs and galaxy mergers.
   
   \item Hints of positive feedback are suggested when we plot $\mathrm{M_{K,nuclear}}$ versus $\mathrm{M_{K,bulge}}$ of the sample. We find that 
   RL--NLSy1s behave in a similar manner as FSRQs (or high luminosity blazars), which might be the result of a similar accretion mode 
   between RL--NLSy1s and FSRQs.
   
 \end{itemize}

\section*{Acknowledgements}
This work was supported by CONACyT research grant 280789 (M\'exico). JK acknowledges financial support from the Academy of Finland, grant 311438.

%%%%%%%%%%%%%%%%%%%%%%%%%%%%%%%%%%%%%%%%%%%%%%%%%%



\bibliographystyle{mnras}
\bibliography{ms_12dec15.bib} % if your bibtex file is called example.bib

\begin{thebibliography}{}
\makeatletter
\relax
\def\mn@urlcharsother{\let\do\@makeother \do\$\do\&\do\#\do\^\do\_\do\%\do\~}
\def\mn@doi{\begingroup\mn@urlcharsother \@ifnextchar [ {\mn@doi@}
  {\mn@doi@[]}}
\def\mn@doi@[#1]#2{\def\@tempa{#1}\ifx\@tempa\@empty \href
  {http://dx.doi.org/#2} {doi:#2}\else \href {http://dx.doi.org/#2} {#1}\fi
  \endgroup}
\def\mn@eprint#1#2{\mn@eprint@#1:#2::\@nil}
\def\mn@eprint@arXiv#1{\href {http://arxiv.org/abs/#1} {{\tt arXiv:#1}}}
\def\mn@eprint@dblp#1{\href {http://dblp.uni-trier.de/rec/bibtex/#1.xml}
  {dblp:#1}}
\def\mn@eprint@#1:#2:#3:#4\@nil{\def\@tempa {#1}\def\@tempb {#2}\def\@tempc
  {#3}\ifx \@tempc \@empty \let \@tempc \@tempb \let \@tempb \@tempa \fi \ifx
  \@tempb \@empty \def\@tempb {arXiv}\fi \@ifundefined
  {mn@eprint@\@tempb}{\@tempb:\@tempc}{\expandafter \expandafter \csname
  mn@eprint@\@tempb\endcsname \expandafter{\@tempc}}}

\bibitem[\protect\citeauthoryear{{Abdo} et~al.,}{{Abdo}
  et~al.}{2009a}]{abdo_2009a}
{Abdo} A.~A.,  et~al., 2009a, \mn@doi [\apj] {10.1088/0004-637X/699/2/976},
  \href {http://adsabs.harvard.edu/abs/2009ApJ...699..976A} {699, 976}

\bibitem[\protect\citeauthoryear{{Abdo} et~al.,}{{Abdo}
  et~al.}{2009b}]{abdo_2009}
{Abdo} A.~A.,  et~al., 2009b, \mn@doi [\apjl] {10.1088/0004-637X/707/2/L142},
  \href {http://adsabs.harvard.edu/abs/2009ApJ...707L.142A} {707, L142}

\bibitem[\protect\citeauthoryear{{Abraham} \& {Merrifield}}{{Abraham} \&
  {Merrifield}}{2000}]{abraham_merrifield_2000}
{Abraham} R.~G.,  {Merrifield} M.~R.,  2000, \mn@doi [\aj] {10.1086/316877},
  \href {http://adsabs.harvard.edu/abs/2000AJ....120.2835A} {120, 2835}

\bibitem[\protect\citeauthoryear{{Athanassoula}}{{Athanassoula}}{1992}]{athanassoula_1992}
{Athanassoula} E.,  1992, \mn@doi [\mnras] {10.1093/mnras/259.2.345}, \href
  {http://adsabs.harvard.edu/abs/1992MNRAS.259..345A} {259, 345}

\bibitem[\protect\citeauthoryear{{Athanassoula}, {Puerari}  \&
  {Bosma}}{{Athanassoula} et~al.}{1997}]{athanassoula_1997}
{Athanassoula} E.,  {Puerari} I.,   {Bosma} A.,  1997, \mn@doi [\mnras]
  {10.1093/mnras/286.2.284}, \href
  {http://adsabs.harvard.edu/abs/1997MNRAS.286..284A} {286, 284}

\bibitem[\protect\citeauthoryear{{Bagchi} et~al.,}{{Bagchi}
  et~al.}{2014}]{Bagchi_2014}
{Bagchi} J.,  et~al., 2014, \mn@doi [\apj] {10.1088/0004-637X/788/2/174}, \href
  {http://adsabs.harvard.edu/abs/2014ApJ...788..174B} {788, 174}

\bibitem[\protect\citeauthoryear{{Bertin} \& {Arnouts}}{{Bertin} \&
  {Arnouts}}{1996}]{bertin_1996}
{Bertin} E.,  {Arnouts} S.,  1996, A\&ASS, \href
  {http://adsabs.harvard.edu/abs/1996A%26AS..117..393B} {117, 393}

\bibitem[\protect\citeauthoryear{{Blecha} \& {Loeb}}{{Blecha} \&
  {Loeb}}{2008}]{blecha_2008}
{Blecha} L.,  {Loeb} A.,  2008, \mn@doi [\mnras]
  {10.1111/j.1365-2966.2008.13790.x}, \href
  {https://ui.adsabs.harvard.edu/abs/2008MNRAS.390.1311B} {390, 1311}

\bibitem[\protect\citeauthoryear{{Blecha}, {Brisken}, {Burke-Spolaor},
  {Civano}, {Comerford}, {Darling}, {Lazio}  \& {Maccarone}}{{Blecha}
  et~al.}{2019}]{blecha_2019}
{Blecha} L.,  {Brisken} W.,  {Burke-Spolaor} S.,  {Civano} F.,  {Comerford} J.,
   {Darling} J.,  {Lazio} T. J.~W.,   {Maccarone} T.~J.,  2019, Astro2020:
  Decadal Survey on Astronomy and Astrophysics, \href
  {https://ui.adsabs.harvard.edu/abs/2019astro2020T.318B} {2020, 318}

\bibitem[\protect\citeauthoryear{{Capelo}, {Volonteri}, {Dotti}, {Bellovary},
  {Mayer}  \& {Governato}}{{Capelo} et~al.}{2015}]{capelo_2015}
{Capelo} P.~R.,  {Volonteri} M.,  {Dotti} M.,  {Bellovary} J.~M.,  {Mayer} L.,
   {Governato} F.,  2015, \mn@doi [\mnras] {10.1093/mnras/stu2500}, \href
  {http://adsabs.harvard.edu/abs/2015MNRAS.447.2123C} {447, 2123}

\bibitem[\protect\citeauthoryear{{Chilingarian} \& {Zolotukhin}}{{Chilingarian}
  \& {Zolotukhin}}{2012}]{chilingarian_2012}
{Chilingarian} I.~V.,  {Zolotukhin} I.~Y.,  2012, \mn@doi [MNRAS]
  {10.1111/j.1365-2966.2011.19837.x}, \href
  {http://adsabs.harvard.edu/abs/2012MNRAS.419.1727C} {419, 1727}

\bibitem[\protect\citeauthoryear{{Chilingarian}, {Melchior}  \&
  {Zolotukhin}}{{Chilingarian} et~al.}{2010}]{chilingarian_2010}
{Chilingarian} I.~V.,  {Melchior} A.-L.,   {Zolotukhin} I.~Y.,  2010, \mn@doi
  [MNRAS] {10.1111/j.1365-2966.2010.16506.x}, \href
  {http://adsabs.harvard.edu/abs/2010MNRAS.405.1409C} {405, 1409}

\bibitem[\protect\citeauthoryear{{Comerford}, {Pooley}, {Barrows}, {Greene},
  {Zakamska}, {Madejski}  \& {Cooper}}{{Comerford}
  et~al.}{2015}]{comerford_2015}
{Comerford} J.~M.,  {Pooley} D.,  {Barrows} R.~S.,  {Greene} J.~E.,  {Zakamska}
  N.~L.,  {Madejski} G.~M.,   {Cooper} M.~C.,  2015, \mn@doi [\apj]
  {10.1088/0004-637X/806/2/219}, \href
  {https://ui.adsabs.harvard.edu/abs/2015ApJ...806..219C} {806, 219}

\bibitem[\protect\citeauthoryear{{D'Ammando} et~al.,}{{D'Ammando}
  et~al.}{2012}]{dammando_2012}
{D'Ammando} F.,  et~al., 2012, \mn@doi [\mnras]
  {10.1111/j.1365-2966.2012.21707.x}, \href
  {http://adsabs.harvard.edu/abs/2012MNRAS.426..317D} {426, 317}

\bibitem[\protect\citeauthoryear{{D'Ammando}, {Orienti}, {Larsson}  \&
  {Giroletti}}{{D'Ammando} et~al.}{2015}]{dammando_2015}
{D'Ammando} F.,  {Orienti} M.,  {Larsson} J.,   {Giroletti} M.,  2015, \mn@doi
  [\mnras] {10.1093/mnras/stv1278}, \href
  {http://adsabs.harvard.edu/abs/2015MNRAS.452..520D} {452, 520}

\bibitem[\protect\citeauthoryear{{D'Ammando}, {Acosta-Pulido}, {Capetti},
  {Baldi}, {Orienti}, {Raiteri}  \& {Ramos Almeida}}{{D'Ammando}
  et~al.}{2018}]{dammando_2018}
{D'Ammando} F.,  {Acosta-Pulido} J.~A.,  {Capetti} A.,  {Baldi} R.~D.,
  {Orienti} M.,  {Raiteri} C.~M.,   {Ramos Almeida} C.,  2018, \mn@doi [\mnras]
  {10.1093/mnrasl/sly072}, \href
  {http://adsabs.harvard.edu/abs/2018MNRAS.478L..66D} {478, L66}

\bibitem[\protect\citeauthoryear{{Di Matteo}, {Croft}, {Springel}  \&
  {Hernquist}}{{Di Matteo} et~al.}{2003}]{dimatteo_2003}
{Di Matteo} T.,  {Croft} R.~A.~C.,  {Springel} V.,   {Hernquist} L.,  2003,
  \mn@doi [\apj] {10.1086/376501}, \href
  {http://adsabs.harvard.edu/abs/2003ApJ...593...56D} {593, 56}

\bibitem[\protect\citeauthoryear{{Ellison}, {Patton}, {Mendel}  \&
  {Scudder}}{{Ellison} et~al.}{2011}]{ellison_2011}
{Ellison} S.~L.,  {Patton} D.~R.,  {Mendel} J.~T.,   {Scudder} J.~M.,  2011,
  \mn@doi [\mnras] {10.1111/j.1365-2966.2011.19624.x}, \href
  {http://adsabs.harvard.edu/abs/2011MNRAS.418.2043E} {418, 2043}

\bibitem[\protect\citeauthoryear{{Ellison}, {Mendel}, {Patton}  \&
  {Scudder}}{{Ellison} et~al.}{2013}]{ellison_2013}
{Ellison} S.~L.,  {Mendel} J.~T.,  {Patton} D.~R.,   {Scudder} J.~M.,  2013,
  \mn@doi [\mnras] {10.1093/mnras/stt1562}, \href
  {http://adsabs.harvard.edu/abs/2013MNRAS.435.3627E} {435, 3627}

\bibitem[\protect\citeauthoryear{{Erwin}}{{Erwin}}{2005}]{erwin_2005}
{Erwin} P.,  2005, \mn@doi [\mnras] {10.1111/j.1365-2966.2005.09560.x}, \href
  {http://adsabs.harvard.edu/abs/2005MNRAS.364..283E} {364, 283}

\bibitem[\protect\citeauthoryear{{Fabian}}{{Fabian}}{2012}]{Fabian_2012}
{Fabian} A.~C.,  2012, \mn@doi [ARA\&A] {10.1146/annurev-astro-081811-125521},
  \href {http://adsabs.harvard.edu/abs/2012ARA%26A..50..455F} {50, 455}

\bibitem[\protect\citeauthoryear{{Ferrarese} \& {Merritt}}{{Ferrarese} \&
  {Merritt}}{2000}]{ferrarese_merrit_2000}
{Ferrarese} L.,  {Merritt} D.,  2000, \mn@doi [ApJL] {10.1086/312838}, \href
  {http://adsabs.harvard.edu/abs/2000ApJ...539L...9F} {539, L9}

\bibitem[\protect\citeauthoryear{{Fisher} \& {Drory}}{{Fisher} \&
  {Drory}}{2008}]{fisher_drory_2008}
{Fisher} D.~B.,  {Drory} N.,  2008, in {Funes} J.~G.,  {Corsini} E.~M.,  eds,
  Astronomical Society of the Pacific Conference Series Vol. 396, Formation and
  Evolution of Galaxy Disks. p.~309

\bibitem[\protect\citeauthoryear{{Foschini}}{{Foschini}}{2011}]{foschini_2011}
{Foschini} L.,  2011, in Narrow-Line Seyfert 1 Galaxies and their Place in the
  Universe. p.~24 (\mn@eprint {arXiv} {1105.0772})

\bibitem[\protect\citeauthoryear{{Foschini} et~al.,}{{Foschini}
  et~al.}{2015}]{foschini_2015}
{Foschini} L.,  et~al., 2015, \mn@doi [\aap] {10.1051/0004-6361/201424972},
  \href {http://adsabs.harvard.edu/abs/2015A%26A...575A..13F} {575, A13}

\bibitem[\protect\citeauthoryear{{Gadotti}}{{Gadotti}}{2009}]{gadotti_2009}
{Gadotti} D.~A.,  2009, \mn@doi [mnras] {10.1111/j.1365-2966.2008.14257.x},
  \href {http://adsabs.harvard.edu/abs/2009MNRAS.393.1531G} {393, 1531}

\bibitem[\protect\citeauthoryear{{Gebhardt} et~al.,}{{Gebhardt}
  et~al.}{2000}]{gebhardt_2000}
{Gebhardt} K.,  et~al., 2000, \mn@doi [ApJL] {10.1086/312840}, \href
  {http://adsabs.harvard.edu/abs/2000ApJ...539L..13G} {539, L13}

\bibitem[\protect\citeauthoryear{{Glass} \& {Moorwood}}{{Glass} \&
  {Moorwood}}{1985}]{glass_1985}
{Glass} I.~S.,  {Moorwood} A.~F.~M.,  1985, \mn@doi [\mnras]
  {10.1093/mnras/214.4.429}, \href
  {http://adsabs.harvard.edu/abs/1985MNRAS.214..429G} {214, 429}

\bibitem[\protect\citeauthoryear{{G{\"u}ltekin} et~al.,}{{G{\"u}ltekin}
  et~al.}{2009}]{Gultekin_2009}
{G{\"u}ltekin} K.,  et~al., 2009, \mn@doi [ApJ] {10.1088/0004-637X/698/1/198},
  \href {http://adsabs.harvard.edu/abs/2009ApJ...698..198G} {698, 198}

\bibitem[\protect\citeauthoryear{{Heckman} \& {Best}}{{Heckman} \&
  {Best}}{2014}]{Heckman_2014}
{Heckman} T.~M.,  {Best} P.~N.,  2014, \mn@doi [ARA\&A]
  {10.1146/annurev-astro-081913-035722}, \href
  {http://adsabs.harvard.edu/abs/2014ARA%26A..52..589H} {52, 589}

\bibitem[\protect\citeauthoryear{{Hopkins}, {Hernquist}, {Hayward}  \&
  {Narayanan}}{{Hopkins} et~al.}{2012}]{hopkins_2012}
{Hopkins} P.~F.,  {Hernquist} L.,  {Hayward} C.~C.,   {Narayanan} D.,  2012,
  \mn@doi [\mnras] {10.1111/j.1365-2966.2012.21449.x}, \href
  {https://ui.adsabs.harvard.edu/abs/2012MNRAS.425.1121H} {425, 1121}

\bibitem[\protect\citeauthoryear{{Hota} et~al.,}{{Hota}
  et~al.}{2011}]{hota_2011}
{Hota} A.,  et~al., 2011, \mn@doi [\mnras] {10.1111/j.1745-3933.2011.01115.x},
  \href {http://adsabs.harvard.edu/abs/2011MNRAS.417L..36H} {417, L36}

\bibitem[\protect\citeauthoryear{{Hyv{\"o}nen}, {Kotilainen}, {Falomo},
  {{\"O}rndahl}  \& {Pursimo}}{{Hyv{\"o}nen} et~al.}{2007}]{Hyvonen_2007}
{Hyv{\"o}nen} T.,  {Kotilainen} J.~K.,  {Falomo} R.,  {{\"O}rndahl} E.,
  {Pursimo} T.,  2007, \mn@doi [A\&A] {10.1051/0004-6361:20078202}, \href
  {http://adsabs.harvard.edu/abs/2007A%26A...476..723H} {476, 723}

\bibitem[\protect\citeauthoryear{{Ishibashi} \& {Fabian}}{{Ishibashi} \&
  {Fabian}}{2012}]{ishibashi_2012}
{Ishibashi} W.,  {Fabian} A.~C.,  2012, \mn@doi [\mnras]
  {10.1111/j.1365-2966.2012.22074.x}, \href
  {https://ui.adsabs.harvard.edu/abs/2012MNRAS.427.2998I} {427, 2998}

\bibitem[\protect\citeauthoryear{{Jogee}, {Kenney}  \& {Smith}}{{Jogee}
  et~al.}{1999}]{jogee_1999}
{Jogee} S.,  {Kenney} J.~D.~P.,   {Smith} B.~J.,  1999, \mn@doi [\apj]
  {10.1086/308021}, \href {http://adsabs.harvard.edu/abs/1999ApJ...526..665J}
  {526, 665}

\bibitem[\protect\citeauthoryear{{Kaspi}, {Smith}, {Netzer}, {Maoz}, {Jannuzi}
  \& {Giveon}}{{Kaspi} et~al.}{2000}]{kaspi_2000}
{Kaspi} S.,  {Smith} P.~S.,  {Netzer} H.,  {Maoz} D.,  {Jannuzi} B.~T.,
  {Giveon} U.,  2000, \mn@doi [\apj] {10.1086/308704}, \href
  {http://adsabs.harvard.edu/abs/2000ApJ...533..631J} {533, 631}

\bibitem[\protect\citeauthoryear{{Kaviraj}, {Shabala}, {Deller}  \&
  {Middelberg}}{{Kaviraj} et~al.}{2015}]{kaviraj_2015}
{Kaviraj} S.,  {Shabala} S.~S.,  {Deller} A.~T.,   {Middelberg} E.,  2015,
  \mn@doi [\mnras] {10.1093/mnras/stv1957}, \href
  {http://adsabs.harvard.edu/abs/2015MNRAS.454.1595K} {454, 1595}

\bibitem[\protect\citeauthoryear{{Komossa}, {Voges}, {Xu}, {Mathur}, {Adorf},
  {Lemson}, {Duschl}  \& {Grupe}}{{Komossa} et~al.}{2006}]{komossa_2006}
{Komossa} S.,  {Voges} W.,  {Xu} D.,  {Mathur} S.,  {Adorf} H.-M.,  {Lemson}
  G.,  {Duschl} W.~J.,   {Grupe} D.,  2006, \mn@doi [\aj] {10.1086/505043},
  \href {http://adsabs.harvard.edu/abs/2006AJ....132..531K} {132, 531}

\bibitem[\protect\citeauthoryear{{Kormendy}}{{Kormendy}}{1977}]{Kormendy_1977}
{Kormendy} J.,  1977, \mn@doi [ApJ] {10.1086/155687}, \href
  {http://adsabs.harvard.edu/abs/1977ApJ...218..333K} {218, 333}

\bibitem[\protect\citeauthoryear{{Koss}, {Mushotzky}, {Treister}, {Veilleux},
  {Vasudevan}  \& {Trippe}}{{Koss} et~al.}{2012}]{koss_2012}
{Koss} M.,  {Mushotzky} R.,  {Treister} E.,  {Veilleux} S.,  {Vasudevan} R.,
  {Trippe} M.,  2012, \mn@doi [\apjl] {10.1088/2041-8205/746/2/L22}, \href
  {http://adsabs.harvard.edu/abs/2012ApJ...746L..22K} {746, L22}

\bibitem[\protect\citeauthoryear{{Kotilainen}, {Falomo}  \&
  {Scarpa}}{{Kotilainen} et~al.}{1998a}]{kotilainen_1998}
{Kotilainen} J.~K.,  {Falomo} R.,   {Scarpa} R.,  1998a, A\&A, \href
  {http://adsabs.harvard.edu/abs/1998A%26A...332..503K} {332, 503}

\bibitem[\protect\citeauthoryear{{Kotilainen}, {Falomo}  \&
  {Scarpa}}{{Kotilainen} et~al.}{1998b}]{Kotilainen_1998bllacs}
{Kotilainen} J.~K.,  {Falomo} R.,   {Scarpa} R.,  1998b, A\&A, \href
  {http://adsabs.harvard.edu/abs/1998A%26A...336..479K} {336, 479}

\bibitem[\protect\citeauthoryear{{Kotilainen}, {Hyv{\"o}nen}  \&
  {Falomo}}{{Kotilainen} et~al.}{2005}]{kotilainen_2005}
{Kotilainen} J.~K.,  {Hyv{\"o}nen} T.,   {Falomo} R.,  2005, \mn@doi [A\&A]
  {10.1051/0004-6361:20042548}, \href
  {http://adsabs.harvard.edu/abs/2005A%26A...440..831K} {440, 831}

\bibitem[\protect\citeauthoryear{{Kotilainen}, {Falomo}, {Labita}, {Treves}  \&
  {Uslenghi}}{{Kotilainen} et~al.}{2007}]{kotilainen_2007}
{Kotilainen} J.~K.,  {Falomo} R.,  {Labita} M.,  {Treves} A.,   {Uslenghi} M.,
  2007, \mn@doi [ApJ] {10.1086/512847}, \href
  {http://adsabs.harvard.edu/abs/2007ApJ...660.1039K} {660, 1039}

\bibitem[\protect\citeauthoryear{{Kotilainen}, {Tavares}, {Olguin-Iglesias},
  {Baes}, {Anorve}, {Chavushyan}  \& {Carrasco}}{{Kotilainen}
  et~al.}{2016}]{kotilainen_2016}
{Kotilainen} J.~K.,  {Tavares} J.~L.,  {Olguin-Iglesias} A.,  {Baes} M.,
  {Anorve} C.,  {Chavushyan} V.,   {Carrasco} L.,  2016, preprint, \href
  {http://adsabs.harvard.edu/abs/2016arXiv160902417K} {} (\mn@eprint {arXiv}
  {1609.02417})

\bibitem[\protect\citeauthoryear{{Laurikainen}, {Salo}  \&
  {Rautiainen}}{{Laurikainen} et~al.}{2002}]{laurikainen_2002}
{Laurikainen} E.,  {Salo} H.,   {Rautiainen} P.,  2002, \mn@doi [\mnras]
  {10.1046/j.1365-8711.2002.05243.x}, \href
  {http://adsabs.harvard.edu/abs/2002MNRAS.331..880L} {331, 880}

\bibitem[\protect\citeauthoryear{{Laurikainen}, {Salo}  \&
  {Buta}}{{Laurikainen} et~al.}{2004}]{laurikainen_2004}
{Laurikainen} E.,  {Salo} H.,   {Buta} R.,  2004, \mn@doi [\apj]
  {10.1086/383462}, \href {http://adsabs.harvard.edu/abs/2004ApJ...607..103L}
  {607, 103}

\bibitem[\protect\citeauthoryear{{Laurikainen}, {Salo}, {Buta}  \&
  {Knapen}}{{Laurikainen} et~al.}{2007}]{laurikainen_2007}
{Laurikainen} E.,  {Salo} H.,  {Buta} R.,   {Knapen} J.~H.,  2007, \mn@doi
  [\mnras] {10.1111/j.1365-2966.2007.12299.x}, \href
  {http://adsabs.harvard.edu/abs/2007MNRAS.381..401L} {381, 401}

\bibitem[\protect\citeauthoryear{{Ledlow}, {Owen}  \& {Keel}}{{Ledlow}
  et~al.}{1998}]{ledlow_1998}
{Ledlow} M.~J.,  {Owen} F.~N.,   {Keel} W.~C.,  1998, \mn@doi [\apj]
  {10.1086/305251}, \href {http://adsabs.harvard.edu/abs/1998ApJ...495..227L}
  {495, 227}

\bibitem[\protect\citeauthoryear{{Le{\'o}n Tavares} et~al.,}{{Le{\'o}n Tavares}
  et~al.}{2014}]{leontavares_2014}
{Le{\'o}n Tavares} J.,  et~al., 2014, \mn@doi [ApJ]
  {10.1088/0004-637X/795/1/58}, \href
  {http://adsabs.harvard.edu/abs/2014ApJ...795...58L} {795, 58}

\bibitem[\protect\citeauthoryear{{Liao}, {Liang}, {Weng}, {Berton}, {Gu}  \&
  {Fan}}{{Liao} et~al.}{2015}]{liao_2015}
{Liao} N.-H.,  {Liang} Y.-F.,  {Weng} S.-S.,  {Berton} M.,  {Gu} M.-F.,   {Fan}
  Y.-Z.,  2015, preprint, \href
  {http://adsabs.harvard.edu/abs/2015arXiv151005584L} {} (\mn@eprint {arXiv}
  {1510.05584})

\bibitem[\protect\citeauthoryear{{Magorrian} et~al.,}{{Magorrian}
  et~al.}{1998}]{Magorrian_1998}
{Magorrian} J.,  et~al., 1998, \mn@doi [AJ] {10.1086/300353}, \href
  {http://adsabs.harvard.edu/abs/1998AJ....115.2285M} {115, 2285}

\bibitem[\protect\citeauthoryear{{Mao} et~al.,}{{Mao} et~al.}{2015}]{mao_2015}
{Mao} M.~Y.,  et~al., 2015, \mn@doi [\mnras] {10.1093/mnras/stu2302}, \href
  {http://adsabs.harvard.edu/abs/2015MNRAS.446.4176M} {446, 4176}

\bibitem[\protect\citeauthoryear{{Mapelli}, {Moore}  \&
  {Bland-Hawthorn}}{{Mapelli} et~al.}{2008}]{mapelli_2008}
{Mapelli} M.,  {Moore} B.,   {Bland-Hawthorn} J.,  2008, \mn@doi [\mnras]
  {10.1111/j.1365-2966.2008.13421.x}, \href
  {http://adsabs.harvard.edu/abs/2008MNRAS.388..697M} {388, 697}

\bibitem[\protect\citeauthoryear{{Mathur}, {Fields}, {Peterson}  \&
  {Grupe}}{{Mathur} et~al.}{2012a}]{mathur_2012}
{Mathur} S.,  {Fields} D.,  {Peterson} B.~M.,   {Grupe} D.,  2012a, \mn@doi
  [\apj] {10.1088/0004-637X/754/2/146}, \href
  {http://adsabs.harvard.edu/abs/2012ApJ...754..146M} {754, 146}

\bibitem[\protect\citeauthoryear{{Mathur}, {Fields}, {Peterson}  \&
  {Grupe}}{{Mathur} et~al.}{2012b}]{marthur_2012}
{Mathur} S.,  {Fields} D.,  {Peterson} B.~M.,   {Grupe} D.,  2012b, \mn@doi
  [\apj] {10.1088/0004-637X/754/2/146}, \href
  {http://adsabs.harvard.edu/abs/2012ApJ...754..146M} {754, 146}

\bibitem[\protect\citeauthoryear{{Mediavilla} \& {Arribas}}{{Mediavilla} \&
  {Arribas}}{1993}]{mediavilla_1993}
{Mediavilla} E.,  {Arribas} S.,  1993, \mn@doi [\nat] {10.1038/365420a0}, \href
  {https://ui.adsabs.harvard.edu/abs/1993Natur.365..420M} {365, 420}

\bibitem[\protect\citeauthoryear{{Men{\'e}ndez-Delmestre}, {Sheth},
  {Schinnerer}, {Jarrett}  \& {Scoville}}{{Men{\'e}ndez-Delmestre}
  et~al.}{2007}]{mendendez_2007}
{Men{\'e}ndez-Delmestre} K.,  {Sheth} K.,  {Schinnerer} E.,  {Jarrett} T.~H.,
  {Scoville} N.~Z.,  2007, \mn@doi [\apj] {10.1086/511025}, \href
  {http://adsabs.harvard.edu/abs/2007ApJ...657..790M} {657, 790}

\bibitem[\protect\citeauthoryear{{Menezes}, {Steiner}  \& {Ricci}}{{Menezes}
  et~al.}{2014}]{menezes_2014}
{Menezes} R.~B.,  {Steiner} J.~E.,   {Ricci} T.~V.,  2014, \mn@doi [\apjl]
  {10.1088/2041-8205/796/1/L13}, \href
  {https://ui.adsabs.harvard.edu/abs/2014ApJ...796L..13M} {796, L13}

\bibitem[\protect\citeauthoryear{{Merritt}}{{Merritt}}{2006}]{merritt_2006}
{Merritt} D.,  2006, \mn@doi [\apj] {10.1086/506139}, \href
  {https://ui.adsabs.harvard.edu/abs/2006ApJ...648..976M} {648, 976}

\bibitem[\protect\citeauthoryear{{Mobasher}, {Sharples}  \& {Ellis}}{{Mobasher}
  et~al.}{1993}]{mobasher_1993}
{Mobasher} B.,  {Sharples} R.~M.,   {Ellis} R.~S.,  1993, \mn@doi [\mnras]
  {10.1093/mnras/263.3.560}, \href
  {https://ui.adsabs.harvard.edu/abs/1993MNRAS.263..560M} {263, 560}

\bibitem[\protect\citeauthoryear{{Moorwood} et~al.,}{{Moorwood}
  et~al.}{1998}]{Isaac}
{Moorwood} A.,  et~al., 1998, The Messenger, \href
  {http://adsabs.harvard.edu/abs/1998Msngr..94....7M} {94, 7}

\bibitem[\protect\citeauthoryear{{Morganti}, {Holt}, {Tadhunter}, {Ramos
  Almeida}, {Dicken}, {Inskip}, {Oosterloo}  \& {Tzioumis}}{{Morganti}
  et~al.}{2011}]{Morganti_2011}
{Morganti} R.,  {Holt} J.,  {Tadhunter} C.,  {Ramos Almeida} C.,  {Dicken} D.,
  {Inskip} K.,  {Oosterloo} T.,   {Tzioumis} T.,  2011, \mn@doi [\aap]
  {10.1051/0004-6361/201117686}, \href
  {http://adsabs.harvard.edu/abs/2011A%26A...535A..97M} {535, A97}

\bibitem[\protect\citeauthoryear{{Moriondo}, {Giovanardi}  \&
  {Hunt}}{{Moriondo} et~al.}{1998}]{moriondo_1998}
{Moriondo} G.,  {Giovanardi} C.,   {Hunt} L.~K.,  1998, \mn@doi [\aaps]
  {10.1051/aas:1998408}, \href
  {http://adsabs.harvard.edu/abs/1998A%26AS..130...81M} {130, 81}

\bibitem[\protect\citeauthoryear{{Olgu{\'{\i}}n-Iglesias}
  et~al.,}{{Olgu{\'{\i}}n-Iglesias} et~al.}{2016}]{olguin_2016}
{Olgu{\'{\i}}n-Iglesias} A.,  et~al., 2016, \mn@doi [\mnras]
  {10.1093/mnras/stw1208}, \href
  {http://adsabs.harvard.edu/abs/2016MNRAS.460.3202O} {460, 3202}

\bibitem[\protect\citeauthoryear{{Olgu{\'{\i}}n-Iglesias}, {Kotilainen},
  {Le{\'o}n Tavares}, {Chavushyan}  \& {A{\~n}orve}}{{Olgu{\'{\i}}n-Iglesias}
  et~al.}{2017}]{olguin_2017}
{Olgu{\'{\i}}n-Iglesias} A.,  {Kotilainen} J.~K.,  {Le{\'o}n Tavares} J.,
  {Chavushyan} V.,   {A{\~n}orve} C.,  2017, \mn@doi [\mnras]
  {10.1093/mnras/stx022}, \href
  {http://adsabs.harvard.edu/abs/2017MNRAS.467.3712O} {467, 3712}

\bibitem[\protect\citeauthoryear{{Orban de Xivry}, {Davies}, {Schartmann},
  {Komossa}, {Marconi}, {Hicks}, {Engel}  \& {Tacconi}}{{Orban de Xivry}
  et~al.}{2011}]{orban_2011}
{Orban de Xivry} G.,  {Davies} R.,  {Schartmann} M.,  {Komossa} S.,  {Marconi}
  A.,  {Hicks} E.,  {Engel} H.,   {Tacconi} L.,  2011, \mn@doi [\mnras]
  {10.1111/j.1365-2966.2011.19439.x}, \href
  {http://adsabs.harvard.edu/abs/2011MNRAS.417.2721O} {417, 2721}

\bibitem[\protect\citeauthoryear{{Osterbrock} \& {Pogge}}{{Osterbrock} \&
  {Pogge}}{1985}]{osterbrok_pogge_1985}
{Osterbrock} D.~E.,  {Pogge} R.~W.,  1985, \mn@doi [\apj] {10.1086/163513},
  \href {http://adsabs.harvard.edu/abs/1985ApJ...297..166O} {297, 166}

\bibitem[\protect\citeauthoryear{{Paliya} \& {Stalin}}{{Paliya} \&
  {Stalin}}{2016}]{paliya_2016}
{Paliya} V.~S.,  {Stalin} C.~S.,  2016, \mn@doi [\apj]
  {10.3847/0004-637X/820/1/52}, \href
  {http://adsabs.harvard.edu/abs/2016ApJ...820...52P} {820, 52}

\bibitem[\protect\citeauthoryear{{Paliya}, {Stalin}, {Kumar}, {Kumar}, {Bhatt},
  {Pandey}  \& {Yadav}}{{Paliya} et~al.}{2013}]{paliya_2013}
{Paliya} V.~S.,  {Stalin} C.~S.,  {Kumar} B.,  {Kumar} B.,  {Bhatt} V.~K.,
  {Pandey} S.~B.,   {Yadav} R.~K.~S.,  2013, \mn@doi [\mnras]
  {10.1093/mnras/sts217}, \href
  {http://adsabs.harvard.edu/abs/2013MNRAS.428.2450P} {428, 2450}

\bibitem[\protect\citeauthoryear{{Paliya}, {Ajello}, {Rakshit}, {Mandal},
  {Stalin}, {Kaur}  \& {Hartmann}}{{Paliya} et~al.}{2018}]{paliya_2018}
{Paliya} V.~S.,  {Ajello} M.,  {Rakshit} S.,  {Mandal} A.~K.,  {Stalin} C.~S.,
  {Kaur} A.,   {Hartmann} D.,  2018, \mn@doi [\apjl]
  {10.3847/2041-8213/aaa5ab}, \href
  {http://adsabs.harvard.edu/abs/2018ApJ...853L...2P} {853, L2}

\bibitem[\protect\citeauthoryear{{Peng}, {Ho}, {Impey}  \& {Rix}}{{Peng}
  et~al.}{2011}]{peng_2011}
{Peng} C.~Y.,  {Ho} L.~C.,  {Impey} C.~D.,   {Rix} H.-W.,  2011, {GALFIT:
  Detailed Structural Decomposition of Galaxy Images}, Astrophysics Source Code
  Library (\mn@eprint {ascl} {1104.010})

\bibitem[\protect\citeauthoryear{{Qu}, {Di Matteo}, {Lehnert}, {van Driel}  \&
  {Jog}}{{Qu} et~al.}{2011}]{qu_2011}
{Qu} Y.,  {Di Matteo} P.,  {Lehnert} M.~D.,  {van Driel} W.,   {Jog} C.~J.,
  2011, \mn@doi [\aap] {10.1051/0004-6361/201116502}, \href
  {http://adsabs.harvard.edu/abs/2011A%26A...535A...5Q} {535, A5}

\bibitem[\protect\citeauthoryear{{Reichard}, {Heckman}, {Rudnick},
  {Brinchmann}, {Kauffmann}  \& {Wild}}{{Reichard}
  et~al.}{2009}]{reichard_2009}
{Reichard} T.~A.,  {Heckman} T.~M.,  {Rudnick} G.,  {Brinchmann} J.,
  {Kauffmann} G.,   {Wild} V.,  2009, \mn@doi [\apj]
  {10.1088/0004-637X/691/2/1005}, \href
  {http://adsabs.harvard.edu/abs/2009ApJ...691.1005R} {691, 1005}

\bibitem[\protect\citeauthoryear{{Scarpa}, {Urry}, {Falomo}, {Pesce}  \&
  {Treves}}{{Scarpa} et~al.}{2000}]{scarpa_2000}
{Scarpa} R.,  {Urry} C.~M.,  {Falomo} R.,  {Pesce} J.~E.,   {Treves} A.,  2000,
  \mn@doi [ApJ] {10.1086/308618}, \href
  {http://adsabs.harvard.edu/abs/2000ApJ...532..740S} {532, 740}

\bibitem[\protect\citeauthoryear{{Seigar} \& {James}}{{Seigar} \&
  {James}}{1998}]{seigar_1998}
{Seigar} M.~S.,  {James} P.~A.,  1998, \mn@doi [\mnras]
  {10.1046/j.1365-8711.1998.01778.x}, \href
  {http://adsabs.harvard.edu/abs/1998MNRAS.299..672S} {299, 672}

\bibitem[\protect\citeauthoryear{{Silk}}{{Silk}}{2013}]{silk_2013}
{Silk} J.,  2013, \mn@doi [\apj] {10.1088/0004-637X/772/2/112}, \href
  {https://ui.adsabs.harvard.edu/abs/2013ApJ...772..112S} {772, 112}

\bibitem[\protect\citeauthoryear{{Silverman} et~al.,}{{Silverman}
  et~al.}{2011}]{silverman_2011}
{Silverman} J.~D.,  et~al., 2011, \mn@doi [\apj] {10.1088/0004-637X/743/1/2},
  \href {http://adsabs.harvard.edu/abs/2011ApJ...743....2S} {743, 2}

\bibitem[\protect\citeauthoryear{{Singh}, {Ishwara-Chandra}, {Sievers},
  {Wadadekar}, {Hilton}  \& {Beelen}}{{Singh} et~al.}{2015}]{singh_2015}
{Singh} V.,  {Ishwara-Chandra} C.~H.,  {Sievers} J.,  {Wadadekar} Y.,  {Hilton}
  M.,   {Beelen} A.,  2015, \mn@doi [\mnras] {10.1093/mnras/stv2071}, \href
  {http://adsabs.harvard.edu/abs/2015MNRAS.454.1556S} {454, 1556}

\bibitem[\protect\citeauthoryear{{Skrutskie} et~al.,}{{Skrutskie}
  et~al.}{2006}]{2mass}
{Skrutskie} M.~F.,  et~al., 2006, \mn@doi [\aj] {10.1086/498708}, \href
  {http://adsabs.harvard.edu/abs/2006AJ....131.1163S} {131, 1163}

\bibitem[\protect\citeauthoryear{{Stickel}, {Padovani}, {Urry}, {Fried}  \&
  {Kuehr}}{{Stickel} et~al.}{1991}]{stickel_1991}
{Stickel} M.,  {Padovani} P.,  {Urry} C.~M.,  {Fried} J.~W.,   {Kuehr} H.,
  1991, \mn@doi [apj] {10.1086/170133}, \href
  {http://adsabs.harvard.edu/abs/1991ApJ...374..431S} {374, 431}

\bibitem[\protect\citeauthoryear{{Sundararajan}, {Khanna}  \&
  {Hughes}}{{Sundararajan} et~al.}{2010}]{sundararajan_2010}
{Sundararajan} P.~A.,  {Khanna} G.,   {Hughes} S.~A.,  2010, \mn@doi [\prd]
  {10.1103/PhysRevD.81.104009}, \href
  {https://ui.adsabs.harvard.edu/abs/2010PhRvD..81j4009S} {81, 104009}

\bibitem[\protect\citeauthoryear{{Tremaine} et~al.,}{{Tremaine}
  et~al.}{2002}]{tremaine_2002}
{Tremaine} S.,  et~al., 2002, \mn@doi [ApJ] {10.1086/341002}, \href
  {http://adsabs.harvard.edu/abs/2002ApJ...574..740T} {574, 740}

\bibitem[\protect\citeauthoryear{{Urry}, {Scarpa}, {O'Dowd}, {Falomo}, {Pesce}
  \& {Treves}}{{Urry} et~al.}{2000}]{urry_2000}
{Urry} C.~M.,  {Scarpa} R.,  {O'Dowd} M.,  {Falomo} R.,  {Pesce} J.~E.,
  {Treves} A.,  2000, \mn@doi [ApJ] {10.1086/308616}, \href
  {http://adsabs.harvard.edu/abs/2000ApJ...532..816U} {532, 816}

\bibitem[\protect\citeauthoryear{{Wozniak}, {Friedli}, {Martinet}, {Martin}  \&
  {Bratschi}}{{Wozniak} et~al.}{1995}]{wozniak_1995}
{Wozniak} H.,  {Friedli} D.,  {Martinet} L.,  {Martin} P.,   {Bratschi} P.,
  1995, \aaps, \href {http://adsabs.harvard.edu/abs/1995A%26AS..111..115W}
  {111, 115}

\bibitem[\protect\citeauthoryear{{Yao}, {Yuan}, {Zhou}, {Komossa}, {Zhang},
  {Qiao}  \& {Liu}}{{Yao} et~al.}{2015}]{yao_2015}
{Yao} S.,  {Yuan} W.,  {Zhou} H.,  {Komossa} S.,  {Zhang} J.,  {Qiao} E.,
  {Liu} B.,  2015, \mn@doi [\mnras] {10.1093/mnrasl/slv119}, \href
  {http://adsabs.harvard.edu/abs/2015MNRAS.454L..16Y} {454, L16}

\makeatother
\end{thebibliography}


%%%%%%%%%%%%%%%%%%%%%%%%%%%%%%%%%%%%%%%%%%%%%%%%%%

%%%%%%%%%%%%%%%%% APPENDICES %%%%%%%%%%%%%%%%%%%%%

\appendix

\section{Simulations}\label{table:sim}
 \begin{table*}
\centering
 \begin{minipage}{140mm}
\resizebox{1.0\textwidth}{!}{%
 \begin{tabular}{@{}cccccccccccccc@{}}
  \hline
&\multicolumn{5}{c}{Simulation parameters}& \multicolumn{6}{c}{Retrieved parameters}& &\\
\multirow{2}{*}{}	&	Seeing 		&	$\mathrm{m_{nuclear}}$	&	$\mathrm{m_{bulge}}$	&	$\mathrm{Re}$		&	$\mathrm{m_{disc}}$	&	&$\mathrm{m_{nuclear}}$	& $\mathrm{m_{bulge}}$	&		$\mathrm{R_e}$		&	n	&	$\mathrm{m_{disc}}$	&	$\mathrm{R_s}$		&\multirow{2}{*}{Model quality}	\\
					&	 (arcsec)	&			&			&	(arcsec)		&		&	&		& 		&		(arcsec)	&		&			&	(arcsec)	&\\ 
\cline{2-6} \cline{8-13}\\
1	&	0.60	&	16.00	&	18.50	&	0.10	&	15.00	&	&	15.90	&	20.75	&	2.13	&	0.06	&	15.09	&	2.83	&	bad	\\
2	&	0.60	&	15.00	&	17.00	&	0.10	&	15.00	&	&	14.85	&	19.63	&	2.09	&	0.19	&	15.09	&	2.87	&	bad	\\
3	&	0.60	&	14.00	&	15.50	&	0.10	&	15.00	&	&	13.79	&	17.71	&	0.40	&	0.06	&	15.09	&	2.79	&	bad	\\
4	&	0.60	&	13.00	&	14.00	&	0.10	&	15.00	&	&	12.66	&	17.92	&	0.21	&	1.84	&	15.09	&	2.75	&	bad	\\
5	&	0.60	&	12.00	&	12.50	&	0.10	&	15.00	&	&	11.64	&	13.54	&	0.25	&	0.05	&	15.09	&	2.81	&	bad	\\
6	&	0.60	&	16.00	&	18.50	&	0.10	&	15.00	&	&	15.90	&	20.75	&	2.13	&	0.06	&	15.09	&	2.83	&	bad	\\
7	&	0.60	&	15.00	&	17.00	&	0.10	&	15.00	&	&	14.85	&	19.63	&	2.09	&	0.19	&	15.09	&	2.87	&	bad	\\
8	&	0.60	&	14.00	&	15.50	&	0.10	&	15.00	&	&	13.79	&	17.71	&	0.40	&	0.06	&	15.09	&	2.79	&	bad	\\
9	&	0.60	&	13.00	&	14.00	&	0.10	&	15.00	&	&	12.66	&	17.92	&	0.21	&	1.84	&	15.09	&	2.75	&	bad	\\
10	&	0.60	&	12.00	&	12.50	&	0.10	&	15.00	&	&	11.64	&	13.54	&	0.25	&	0.05	&	15.09	&	2.81	&	bad	\\
11	&	0.60	&	16.00	&	18.50	&	0.56	&	18.50	&	&	15.99	&	18.84	&	0.56	&	0.54	&	15.09	&	2.79	&	good	\\
12	&	0.60	&	15.00	&	17.00	&	0.56	&	15.00	&	&	15.00	&	17.09	&	0.56	&	0.88	&	15.09	&	2.78	&	good	\\
13	&	0.60	&	14.00	&	15.50	&	0.56	&	15.00	&	&	14.00	&	15.52	&	0.56	&	0.97	&	15.09	&	2.78	&	good	\\
14	&	0.60	&	13.00	&	14.00	&	0.56	&	15.00	&	&	13.00	&	14.01	&	0.56	&	0.99	&	15.09	&	2.78	&	good	\\
15	&	0.60	&	12.00	&	12.50	&	0.56	&	15.00	&	&	12.00	&	12.50	&	0.56	&	1.00	&	15.09	&	2.78	&	good	\\
16	&	0.60	&	16.00	&	18.50	&	0.56	&	18.50	&	&	15.99	&	18.84	&	0.56	&	0.54	&	15.09	&	2.79	&	good	\\
17	&	0.60	&	15.00	&	17.00	&	0.56	&	15.00	&	&	15.00	&	17.09	&	0.56	&	0.88	&	15.09	&	2.78	&	good	\\
18	&	0.60	&	14.00	&	15.50	&	0.56	&	15.00	&	&	14.00	&	15.52	&	0.56	&	0.97	&	15.09	&	2.78	&	good	\\
19	&	0.60	&	13.00	&	14.00	&	0.56	&	15.00	&	&	13.00	&	14.01	&	0.56	&	0.99	&	15.09	&	2.78	&	good	\\
20	&	0.60	&	12.00	&	12.50	&	0.56	&	15.00	&	&	12.00	&	12.50	&	0.56	&	1.00	&	15.09	&	2.78	&	good	\\
21	&	0.60	&	16.00	&	18.50	&	1.20	&	15.00	&	&	15.99	&	18.16	&	**	&	2.13	&	15.09	&	2.61	&	bad	\\
22	&	0.60	&	15.00	&	17.00	&	1.20	&	15.00	&	&	14.95	&	16.29	&	**	&	**	&	15.06	&	2.11	&	bad	\\
23	&	0.60	&	14.00	&	15.50	&	1.20	&	15.00	&	&	14.00	&	15.50	&	1.20	&	0.98	&	15.07	&	2.74	&	good	\\
24	&	0.60	&	13.00	&	14.00	&	1.20	&	15.00	&	&	13.00	&	14.02	&	1.20	&	0.99	&	15.06	&	2.71	&	good	\\
25	&	0.60	&	12.00	&	12.50	&	1.20	&	15.00	&	&	12.00	&	12.51	&	1.20	&	1.00	&	15.05	&	2.69	&	good	\\
26	&	0.60	&	16.00	&	18.50	&	2.30	&	15.00	&	&	16.00	&	**	&	**	&	0.08	&	15.06	&	2.71	&	bad	\\
27	&	0.60	&	15.00	&	17.00	&	2.30	&	15.00	&	&	15.00	&	15.43	&	3.42	&	0.99	&	15.92	&	3.87	&	bad	\\
28	&	0.60	&	14.00	&	15.50	&	2.30	&	15.00	&	&	14.00	&	15.01	&	3.03	&	1.12	&	16.30	&	5.15	&	bad	\\
29	&	0.60	&	13.00	&	14.00	&	2.30	&	15.00	&	&	13.00	&	13.68	&	2.65	&	1.16	&	**	&	**	&	good	\\
30	&	0.60	&	12.00	&	12.50	&	2.30	&	15.00	&	&	12.00	&	12.41	&	2.39	&	1.06	&	**	&	**	&	good	\\
31	&	0.60	&	16.00	&	18.50	&	3.40	&	15.00	&	&	15.99	&	**	&	**	&	**	&	15.10	&	2.58	&	bad	\\
32	&	0.60	&	15.00	&	17.00	&	3.40	&	15.00	&	&	14.98	&	**	&	**	&	**	&	15.10	&	2.13	&	bad	\\
33	&	0.60	&	14.00	&	15.50	&	3.40	&	15.00	&	&	14.00	&	15.26	&	4.95	&	0.67	&	15.38	&	0.69	&	bad	\\
34	&	0.60	&	13.00	&	14.00	&	3.40	&	15.00	&	&	13.00	&	15.26	&	4.96	&	0.67	&	13.97	&	0.65	&	bad	\\
35	&	0.60	&	12.00	&	12.50	&	3.40	&	15.00	&	&	12.00	&	12.42	&	2.07	&	1.19	&	**	&	**	&	good	\\
36	&	0.60	&	16.00	&	18.50	&	4.50	&	15.00	&	&	16.00	&	**	&	**	&	**	&	15.03	&	2.81	&	bad	\\
37	&	0.60	&	15.00	&	17.00	&	4.50	&	15.00	&	&	15.00	&	**	&	**	&	**	&	14.91	&	2.79	&	bad	\\
38	&	0.60	&	14.00	&	15.50	&	4.50	&	15.00	&	&	14.00	&	15.12	&	3.91	&	0.86	&	15.43	&	3.54	&	good	\\
39	&	0.60	&	13.00	&	14.00	&	4.50	&	15.00	&	&	13.00	&	13.94	&	4.13	&	0.93	&	15.25	&	3.92	&	good	\\
40	&	0.60	&	12.00	&	12.50	&	4.50	&	15.00	&	&	12.00	&	12.59	&	4.62	&	0.99	&	14.45	&	2.34	&	good	\\
41	&	0.70	&	16.00	&	18.50	&	1.20	&	15.00	&	&	15.99	&	19.21	&	0.00	&	15.28	&	15.09	&	2.62	&	bad	\\
42	&	0.70	&	15.00	&	17.00	&	1.20	&	15.00	&	&	15.00	&	17.50	&	1.32	&	1.02	&	15.09	&	2.80	&	good	\\
43	&	0.70	&	14.00	&	15.50	&	1.20	&	15.00	&	&	14.02	&	15.04	&	2.42	&	2.37	&	16.33	&	3.82	&	good	\\
44	&	0.70	&	13.00	&	14.00	&	1.20	&	15.00	&	&	13.00	&	14.02	&	1.21	&	0.99	&	15.07	&	2.70	&	good	\\
45	&	0.70	&	12.00	&	12.50	&	1.20	&	15.00	&	&	12.00	&	12.51	&	1.21	&	1.00	&	15.06	&	2.68	&	good	\\
46	&	0.80	&	16.00	&	18.50	&	2.30	&	15.00	&	&	16.00	&	16.43	&	70.00	&	0.04	&	15.10	&	2.62	&	bad	\\
47	&	0.80	&	15.00	&	17.00	&	2.30	&	15.00	&	&	14.99	&	16.51	&	2.72	&	0.92	&	15.25	&	2.95	&	bad	\\
\hline
\end{tabular}}\\
\caption{Parameters of the sample of simulated galaxies (columns 2-6) and the retrieved parameters of the simulated galaxies when they are modeled in an identical way to the real galaxies in our sample (columns 7-12).}\label{table:sim}
\scriptsize { Column (1) gives the simulation number; 
(2) the simulation seeing; 
(3) and (4) the simulated nuclear and bulge magnitudes; 
(5) the simulated bulge effective radius;
(6) the simulated disc magnitude;
(7) the modeled nuclear magnitude; 
(8) the modeled bulge magnitude; 
(9) the modeled bulge effective radius;
(10) the modeled bulge S\'ersic index
(11) the modeled disc magnitude;
(12) the modeled disc scale length;
(13) the quality with which the simulation was modeled.\\ 
The '**' symbol shows a physically improbable parameter.\\ 
All the simulated galaxies have bulges with S\'ersic indexes (n=1) and discs with scale lengths ($\mathrm{R_s=3.0}$ arcsec). 
\\}
\end{minipage}
\end{table*}

 \begin{table*}
\centering
 \begin{minipage}{140mm}
\resizebox{1.0\textwidth}{!}{%
 \begin{tabular}{@{}cccccccccccccc@{}}
  \hline
&\multicolumn{5}{c}{Simulation parameters}& \multicolumn{6}{c}{Retrieved parameters}& &\\
\multirow{2}{*}{}	&	Seeing 		&	$\mathrm{m_{nuclear}}$	&	$\mathrm{m_{bulge}}$	&	$\mathrm{Re}$		&	$\mathrm{m_{disc}}$	&	&$\mathrm{m_{nuclear}}$	& $\mathrm{m_{bulge}}$	&		$\mathrm{R_e}$		&	n	&	$\mathrm{m_{disc}}$	&	$\mathrm{R_s}$		&\multirow{2}{*}{Model quality}	\\
					&	 (arcsec)	&			&			&	(arcsec)		&		&	&		& 		&		(arcsec)	&		&			&	(arcsec)	&\\ 
\cline{2-6} \cline{8-13}\\
48	&	0.80	&	14.00	&	15.50	&	2.30	&	15.00	&	&	14.00	&	15.02	&	3.02	&	1.11	&	16.31	&	4.94	&	bad	\\
49	&	0.80	&	13.00	&	14.00	&	2.30	&	15.00	&	&	13.00	&	13.68	&	2.65	&	1.16	&	26.10	&	0.01	&	good	\\
50	&	0.80	&	12.00	&	12.50	&	2.30	&	15.00	&	&	12.00	&	12.41	&	2.39	&	1.06	&	**	&	**	&	good	\\
51	&	0.90	&	16.00	&	18.50	&	3.40	&	15.00	&	&	16.02	&	25.61	&	**	&	6.35	&	15.10	&	2.55	&	bad	\\
52	&	0.90	&	15.00	&	17.00	&	3.40	&	15.00	&	&	15.00	&	16.69	&	**	&	**	&	14.96	&	2.58	&	bad	\\
53	&	0.90	&	14.00	&	15.50	&	3.40	&	15.00	&	&	14.00	&	15.68	&	5.29	&	1.20	&	14.95	&	2.22	&	good	\\
54	&	0.90	&	13.00	&	14.00	&	3.40	&	15.00	&	&	13.00	&	13.78	&	3.46	&	0.97	&	15.99	&	4.15	&	good	\\
55	&	0.90	&	12.00	&	12.50	&	3.40	&	15.00	&	&	12.00	&	12.41	&	3.47	&	1.01	&	17.07	&	**	&	good	\\
56	&	1.00	&	16.00	&	18.50	&	0.10	&	15.00	&	&	15.89	&	17.28	&	3.30	&	0.70	&	15.24	&	3.02	&	bad	\\
57	&	1.00	&	15.00	&	17.00	&	0.10	&	15.00	&	&	14.99	&	17.00	&	1.14	&	**	&	15.09	&	2.87	&	bad	\\
58	&	1.00	&	14.00	&	15.50	&	0.10	&	15.00	&	&	13.77	&	18.46	&	**	&	**	&	15.09	&	2.78	&	bad	\\
59	&	1.00	&	13.00	&	14.00	&	0.10	&	15.00	&	&	12.72	&	15.22	&	**	&	**	&	15.12	&	2.94	&	bad	\\
60	&	1.00	&	12.00	&	12.50	&	0.10	&	15.00	&	&	11.60	&	13.64	&	**	&	**	&	15.10	&	2.82	&	bad	\\
61	&	1.00	&	16.00	&	18.50	&	0.56	&	18.50	&	&	16.03	&	17.52	&	1.77	&	**	&	15.11	&	2.85	&	bad	\\
62	&	1.00	&	15.00	&	17.00	&	0.56	&	15.00	&	&	15.00	&	17.02	&	0.54	&	1.09	&	15.09	&	2.79	&	good	\\
63	&	1.00	&	14.00	&	15.50	&	0.56	&	15.00	&	&	14.00	&	15.51	&	0.55	&	1.00	&	15.09	&	2.78	&	good	\\
64	&	1.00	&	13.00	&	14.00	&	0.56	&	15.00	&	&	13.00	&	14.00	&	0.56	&	1.00	&	15.09	&	2.78	&	good	\\
65	&	1.00	&	12.00	&	12.50	&	0.56	&	15.00	&	&	12.00	&	12.50	&	0.56	&	1.00	&	15.09	&	2.78	&	good	\\
66	&	1.00	&	16.00	&	18.50	&	1.20	&	15.00	&	&	16.21	&	16.87	&	**	&	**	&	15.08	&	2.63	&	bad	\\
67	&	1.00	&	15.00	&	17.00	&	1.20	&	15.00	&	&	14.98	&	16.24	&	**	&	0.99	&	15.06	&	2.25	&	bad	\\
68	&	1.00	&	14.00	&	15.50	&	1.20	&	15.00	&	&	14.00	&	15.57	&	1.20	&	0.94	&	15.07	&	2.72	&	good	\\
69	&	1.00	&	13.00	&	14.00	&	1.20	&	15.00	&	&	13.00	&	14.03	&	1.20	&	0.98	&	15.05	&	2.69	&	good	\\
70	&	1.00	&	12.00	&	12.50	&	1.20	&	15.00	&	&	12.00	&	12.51	&	1.20	&	0.99	&	15.04	&	2.66	&	good	\\
71	&	1.00	&	16.00	&	18.50	&	3.40	&	15.00	&	&	16.00	&	**	&	**	&	**	&	15.07	&	2.63	&	bad	\\
72	&	1.00	&	15.00	&	17.00	&	3.40	&	15.00	&	&	14.99	&	15.99	&	3.52	&	0.80	&	15.43	&	3.17	&	bad	\\
73	&	1.00	&	14.00	&	15.50	&	3.40	&	15.00	&	&	14.00	&	17.12	&	**	&	0.12	&	14.62	&	2.29	&	bad	\\
74	&	1.00	&	13.00	&	14.00	&	3.40	&	15.00	&	&	13.01	&	13.67	&	3.54	&	1.03	&	15.32	&	**	&	good	\\
75	&	1.00	&	12.00	&	12.50	&	3.40	&	15.00	&	&	12.00	&	12.41	&	3.47	&	1.01	&	16.25	&	**	&	good	\\
76	&	1.00	&	16.00	&	18.50	&	4.50	&	15.00	&	&	16.00	&	28.53	&	**	&	0.41	&	15.04	&	2.69	&	bad	\\
77	&	1.00	&	15.00	&	17.00	&	4.50	&	15.00	&	&	15.00	&	19.53	&	**	&	1.16	&	14.92	&	2.79	&	bad	\\
78	&	1.00	&	14.00	&	15.50	&	4.50	&	15.00	&	&	14.00	&	15.87	&	3.63	&	0.78	&	14.90	&	3.06	&	good	\\
79	&	1.00	&	13.00	&	14.00	&	4.50	&	15.00	&	&	13.00	&	14.20	&	4.84	&	0.98	&	14.69	&	2.43	&	good	\\
80	&	1.00	&	12.00	&	12.50	&	4.50	&	15.00	&	&	12.00	&	12.41	&	4.50	&	0.99	&	**	&	**	&	good	\\
81	&	1.10	&	14.00	&	13.20	&	0.10	&	15.00	&	&	13.74	&	13.35	&	0.11	&	0.64	&	15.09	&	2.79	&	good	\\
82	&	1.10	&	14.00	&	14.00	&	0.10	&	15.00	&	&	13.55	&	14.62	&	**	&	**	&	15.11	&	2.94	&	bad	\\
83	&	1.10	&	14.00	&	15.00	&	0.10	&	15.00	&	&	13.68	&	16.95	&	0.12	&	**	&	15.11	&	2.88	&	bad	\\
84	&	1.10	&	14.00	&	15.50	&	0.10	&	15.00	&	&	13.79	&	17.57	&	0.50	&	**	&	15.09	&	2.78	&	bad	\\
85	&	1.10	&	14.00	&	14.50	&	0.10	&	15.00	&	&	13.59	&	15.91	&	0.13	&	4.11	&	15.10	&	2.79	&	bad	\\
86	&	1.10	&	14.00	&	13.20	&	0.40	&	15.00	&	&	13.95	&	13.23	&	0.40	&	0.95	&	15.09	&	2.75	&	good	\\
87	&	1.10	&	14.00	&	14.00	&	0.40	&	15.00	&	&	13.95	&	14.05	&	0.40	&	0.91	&	15.09	&	2.76	&	good	\\
88	&	1.10	&	14.00	&	15.00	&	0.40	&	15.00	&	&	13.96	&	15.11	&	0.40	&	0.81	&	15.09	&	2.76	&	good	\\
89	&	1.10	&	14.00	&	13.00	&	0.40	&	15.00	&	&	13.94	&	13.02	&	0.40	&	0.96	&	15.09	&	2.75	&	good	\\
90	&	1.10	&	14.00	&	14.50	&	0.40	&	15.00	&	&	13.96	&	14.58	&	0.40	&	0.87	&	15.09	&	2.76	&	good	\\
\hline
\end{tabular}}\\
\caption*{Table B1: Continued...}
\scriptsize { Column (1) gives the simulation number; 
(2) the simulation seeing; 
(3) and (4) the simulated nuclear and bulge magnitudes; 
(5) the simulated bulge effective radius;
(6) the simulated disc magnitude;
(7) the modeled nuclear magnitude; 
(8) the modeled bulge magnitude; 
(9) the modeled bulge effective radius;
(10) the modeled bulge S\'ersic index
(11) the modeled disc magnitude;
(12) the modeled disc scale length;
(13) the quality with which the simulation was modeled.\\ 
The '**' symbol shows a physically improbable parameter.\\ 
All the simulated galaxies have bulges with S\'ersic indexes (n=1) and discs with scale lengths ($\mathrm{R_s=3.0}$ arcsec). 
\\}
\end{minipage}
\end{table*}

\clearpage
\section{Morphological analysis}
\subsection{Models}\label{appendix_A}
% %%%%%%%%%%%%%%%%%%%%%%%%%%%%%%%%

 \begin{landscape}
    \begin{figure}
 \caption{Surface brightness profile decomposition of the host galaxies in the sample. 
 Top left: Shows the observed image. 
 Top middle: Shows the image of the model used to describe the surface brightness distribution.
 Top right: Shows the residual image. 
 Middle: Radial profile of the surface brightness distribution. The symbols and the main parameters 
 of the models are explained in the plots. For the complete Appendix see the online material. 
 Bottom: Residuals.} \label{tab:stimuli}
 \ContinuedFloat
 \caption{} \label{tab:stimuli}
   \begin{tabular}{c c} 
   \hspace*{-2.2cm}IRAS 11598-0112 (J--band) & IRAS 11598-0112 (K--band)\\
       \hspace*{-2.2cm}\includegraphics[width=4in]{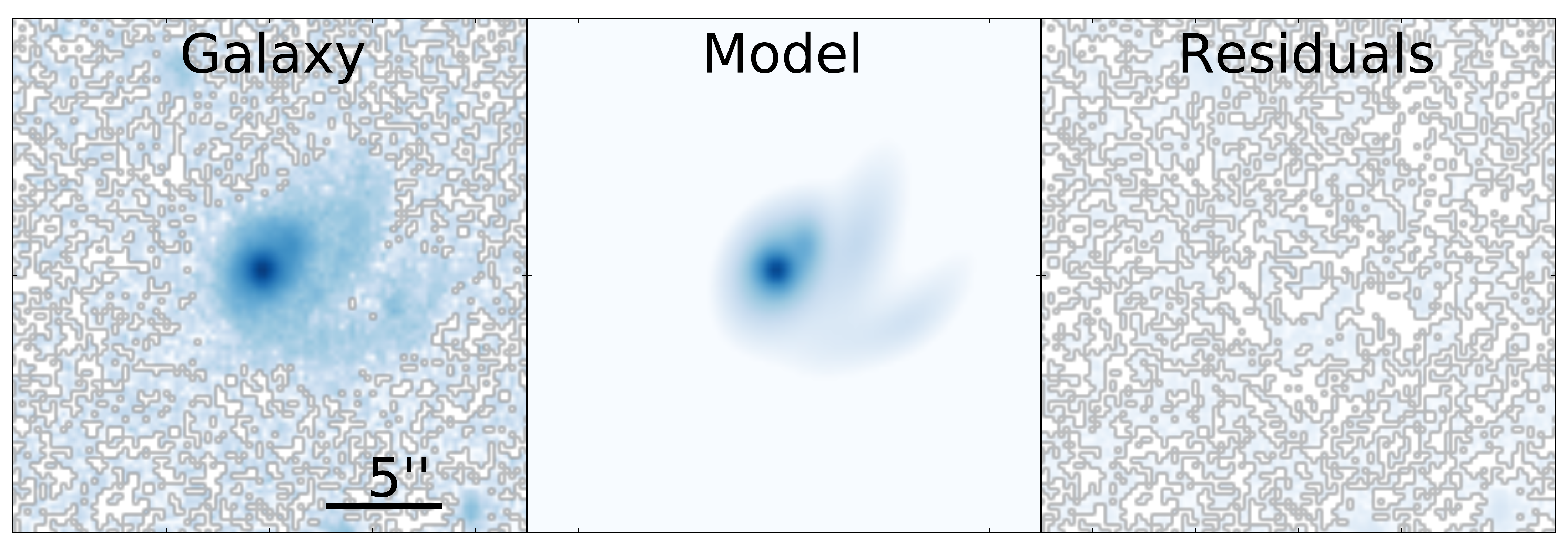} &  \hspace*{.2cm}\includegraphics[width=4in]{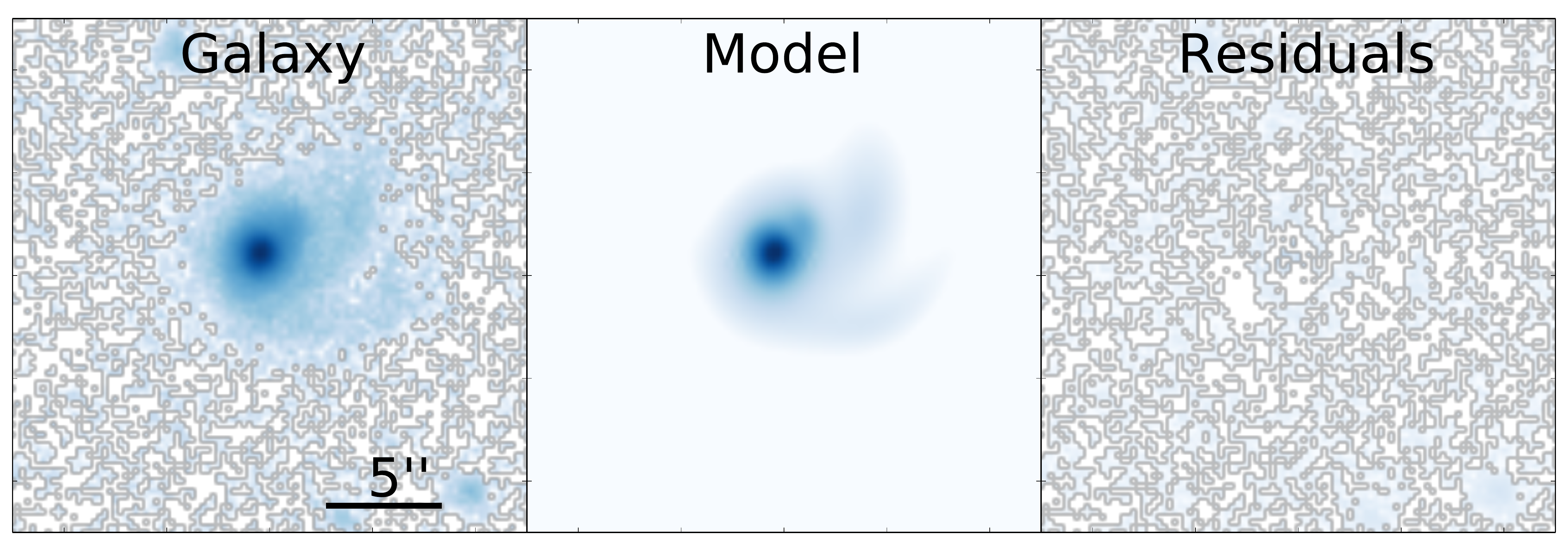}\\ 
       \hspace*{-2.2cm}\includegraphics[width=5.3in]{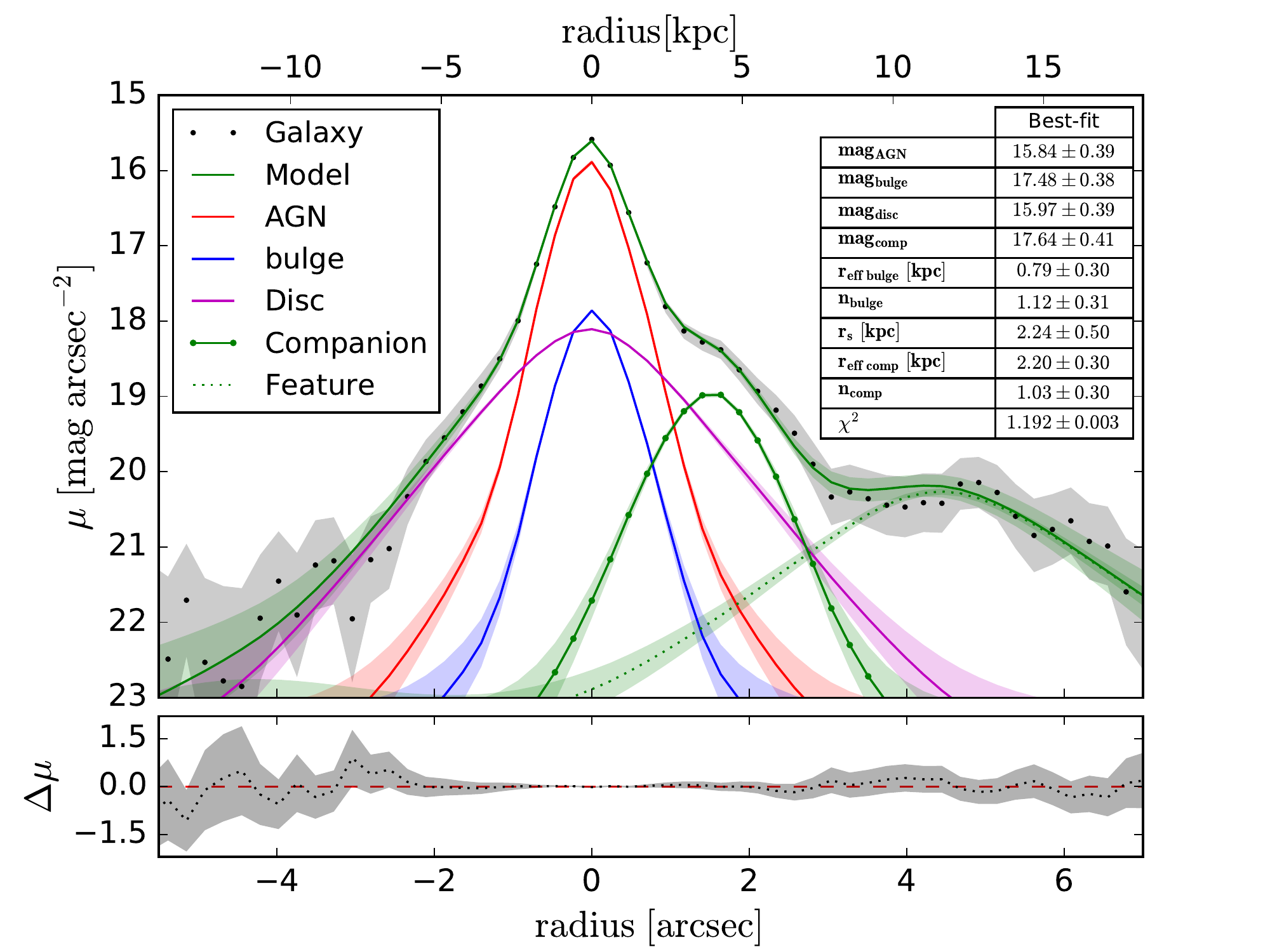} & \includegraphics[width=5.3in]{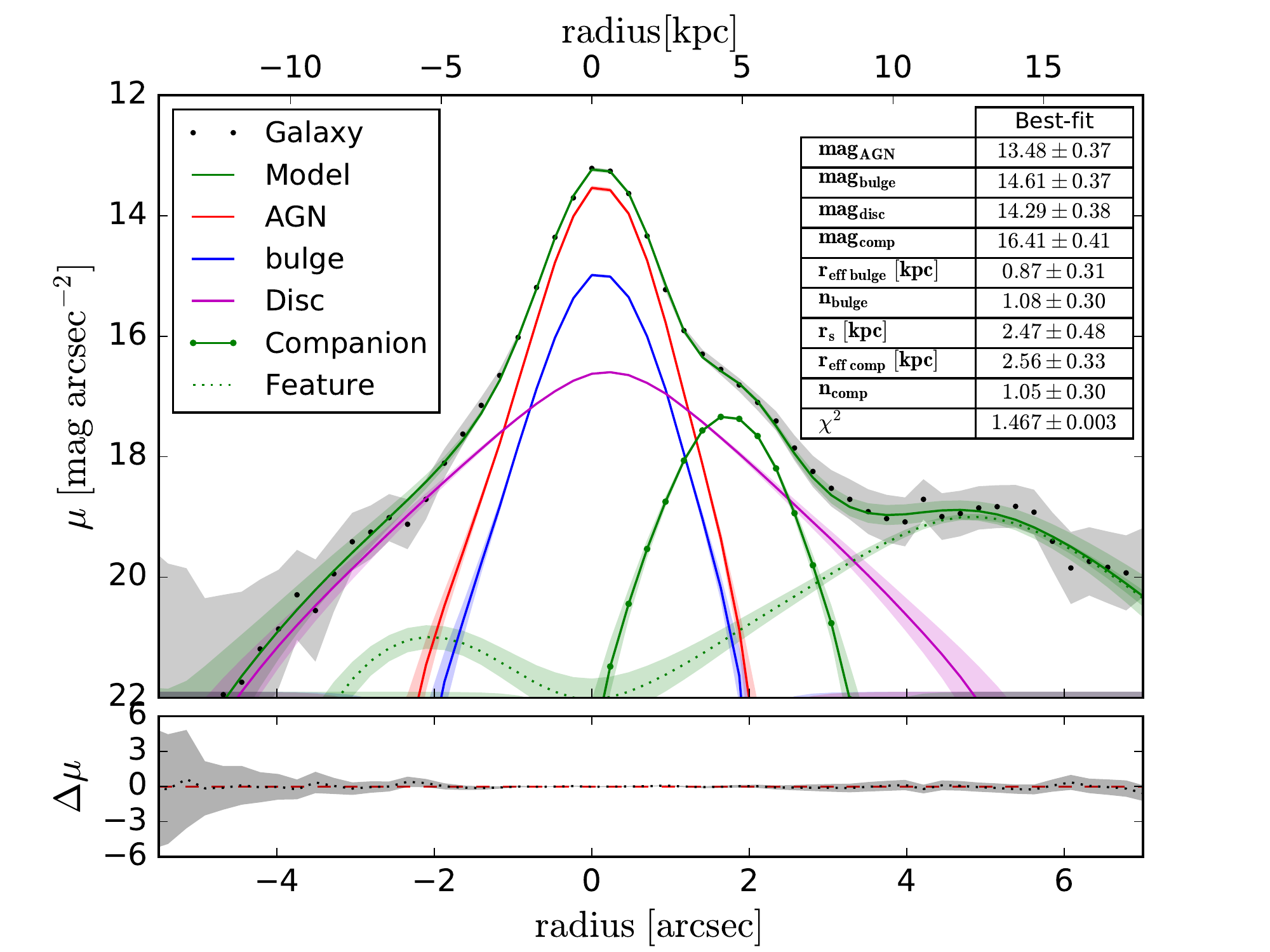}\\
   \end{tabular}
 \end{figure}
 \end{landscape}

 \subsection{Bar test}\label{appendix_B}
\begin{figure*}
\caption{ Radial variation of ellipticity $\epsilon$ (blue circles) and of position angle PA (red squares) derived by ellipse fitting to 
the galaxy isophotes. Only the host galaxies that fulfill the criteria to identify bars 
 are shown (see Section \ref{section:bars}). For the complete Appendix see the online material.} \label{bar_test}
  \begin{tabular}{c} 
  IRAS 11598$-$0112 (J--band) \\
      \hspace*{1.2cm}\includegraphics[width=5in]{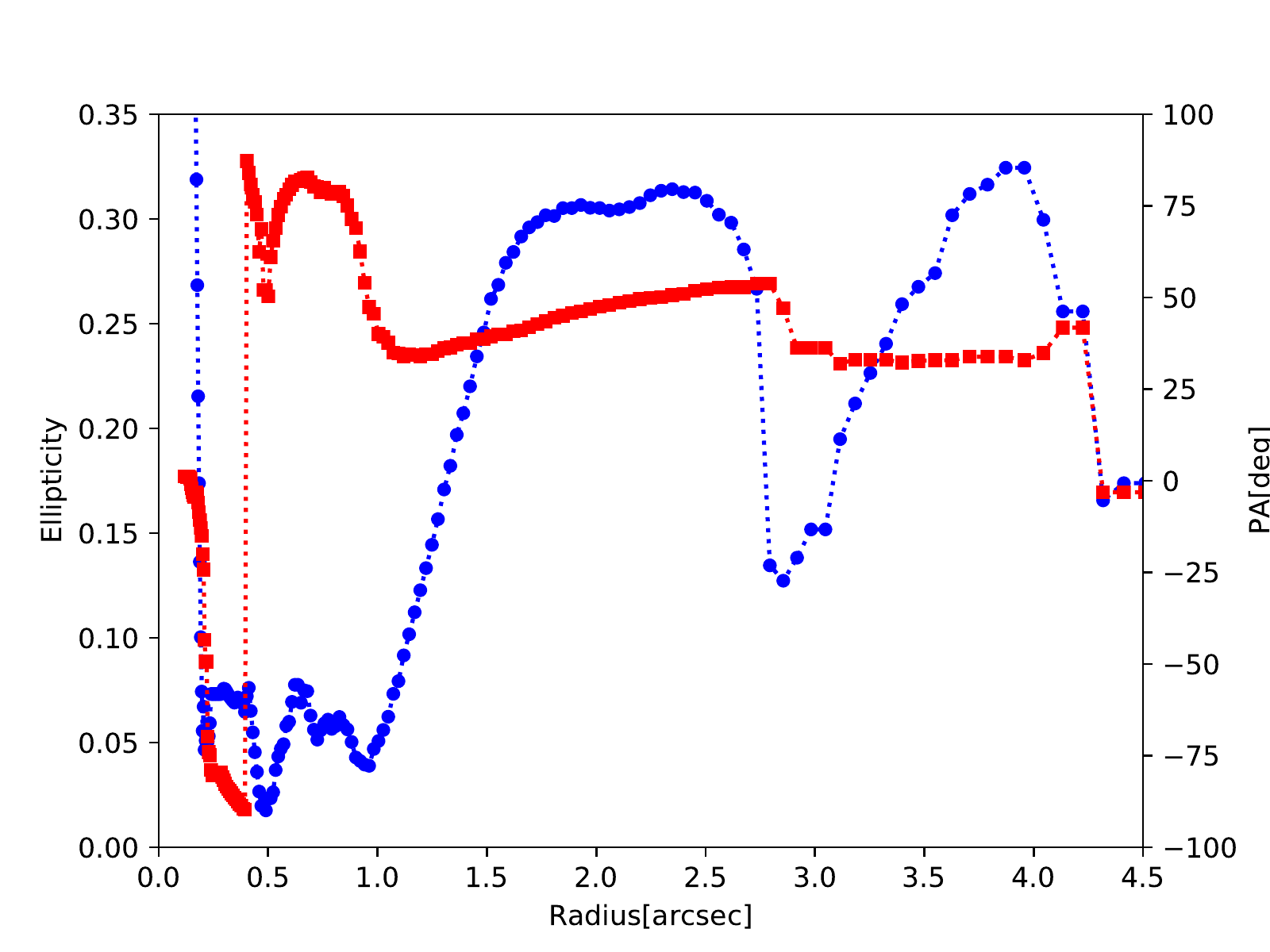}\\ 
       IRAS 11598$-$0112 (K--band)\\
      \hspace*{1.2cm}\includegraphics[width=5.in]{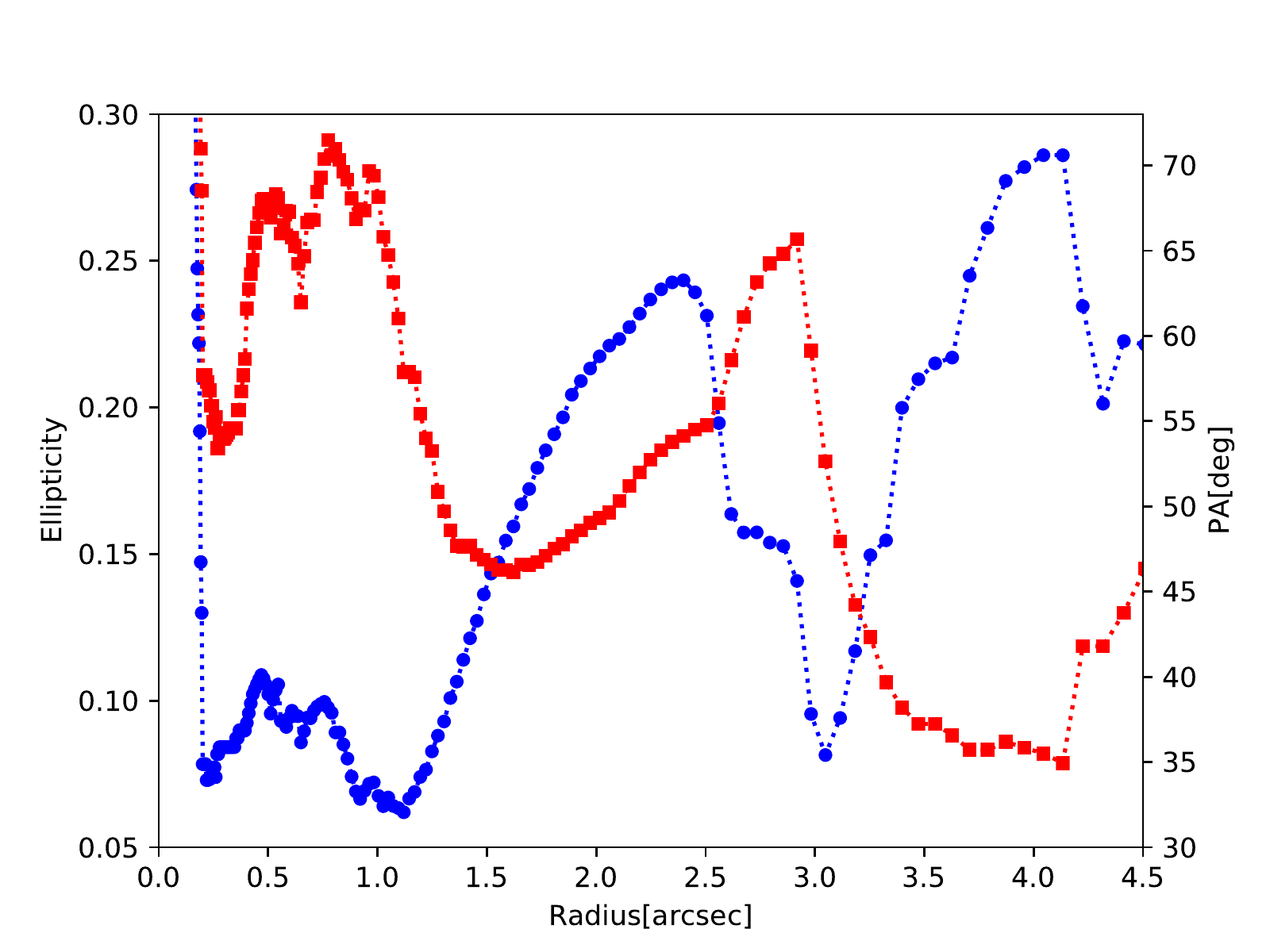}\\
  \end{tabular}
\end{figure*}

%%%%%%%%%%%%%%%%%%%%%%

% Don't change these lines
\bsp	% typesetting comment
\label{lastpage}
\end{document}